\documentclass[journal]{IEEEtran}
\ifCLASSINFOpdf
\usepackage[pdftex]{graphicx}
\else
\fi
\usepackage{amsmath}
\usepackage{amsfonts}
\usepackage{amssymb}
\usepackage{psfrag}
\usepackage{cite}
\usepackage{xcolor}
\usepackage{multirow}
\usepackage[linesnumbered,lined,commentsnumbered]{algorithm2e}
\include{IEEEtran.cls}
\newtheorem{theorem}{Theorem}
\newtheorem{lemma}{Lemma}
\newtheorem{corollary}{Corollary}

\newtheorem{problem}{Problem}
\newtheorem{example}{Example}
\newtheorem{property}{Property}
\newtheorem{algorithm_c}{Algorithm}

\newtheorem{definition}{Definition}
\newcommand{\QED}{{\rm $\blacksquare$}}
\begin{document}
\markboth{IEEE JOURNAL OF SELECTED TOPICS IN SIGNAL PROCESSING, VOL. X, NO. X, Month 2016}{}

\title{Quadrature Amplitude Modulation Division for Multiuser MISO Broadcast Channels }
{\author{Zheng Dong, Yan-Yu Zhang, Jian-Kang Zhang, \emph{Senior Member, IEEE} and Xiang-Chuan Gao
	\thanks{The work of J.-K. Zhang is funded in part by NSERC. Z. Dong would like to thank the support of the China Scholarship Council (CSC) and X.-C. Gao is funded by China National 863 Project (2014AA01A705), Outstanding Young Talent Research Fund of ZZU (1521318003). }
	\thanks{Z. Dong and J.-K. Zhang are with the Department of Electrical and Computer Engineering, McMaster University, Hamilton, Ontario,  Canada. (emails: dongz3@mcmaster.ca, jkzhang@mail.ece.mcmaster.ca.)
	}
	\thanks{Y.-Y. Zhang is with National Digital
Switching System Engineering and Technological
Research Center (NDSC), Zhengzhou (450000), China (email: yyzhang.xinda@gmail.com)}
	\thanks{X.-C. Gao is with the School of Information
		and Engineering, Zhengzhou University, Zhengzhou (450001), China (e-mail:
		iexcgao@zzu.edu.cn)	}}
\maketitle

\begin{abstract}
This paper considers a discrete-time multiuser multiple-input single-output (MISO) Gaussian broadcast channel~(BC), in which channel state information (CSI) is available at both the transmitter and the receivers. The flexible and explicit design of a uniquely decomposable constellation group (UDCG) is provided based on pulse amplitude modulation (PAM) and rectangular quadrature amplitude modulation (QAM) constellations. With this, a modulation division (MD) transmission scheme is developed for the MISO BC.  The proposed MD scheme enables each receiver to uniquely and efficiently detect their desired signals from the superposition of mutually interfering cochannel signals in the absence of noise. In our design, the optimal transmitter beamforming problem is solved in a closed-form for two-user MISO BC using max-min fairness as a design criterion. Then, for a general case with more than two receivers, we develop a user-grouping-based beamforming scheme, where the grouping method, beamforming vector design and power allocation problems are addressed by using weighted max-min fairness. It is shown that our proposed approach has a lower probability of error compared with the zero-forcing (ZF) method when the Hermitian angle between the two channel vectors is small in a two-user case. In addition, simulation results also reveal that for the general channel model with more than two users, our user-grouping-based scheme significantly outperforms the ZF, time division (TD), minimum mean-square error (MMSE) and signal-to-leakage-and-noise ratio (SLNR) based techniques in moderate and high SNR regimes when the number of users approaches to the number of base station (BS) antennas and it degrades into the ZF scheme when the number of users is far less than the number of BS antennas in Rayleigh fading channels.
\end{abstract}


\section{Introduction}
\IEEEPARstart{M}{ultiuser}  systems in a downlink, also known as broadcast channels, have long been the main building block of modern wireless communication systems such as cellular system, telephone conference and digital TV broadcasting. 
In this paper, we concentrate on MISO BC, where one multi-antenna access point serves several single-antenna receivers at the same time. Hence, the design of the receiver can be significantly simplified, since it has only one antenna and the system still enjoys the multi-antenna diversity and multiuser diversity.
BC was first introduced by Cover~\cite{Cover72}, who demonstrated the idea of superposition coding for both binary-symmetric and Gaussian BC. 
Since then, great efforts have been devoted to obtaining the capacity regions of different BCs and to seeking for the optimal transmission strategies under various constraints. 
Although the capacity region for the general discrete memoryless broadcast channel (DM-BC) is still unknown, much progress has been made since~\cite{Cover72}.  In particular, the achievability and converse of the capacity region for the degraded DM-BC were proved by Bergmans~\cite{Bergmans73} and Gallager~\cite{Gallager74}, respectively. Surveys of the literature on the BC can be found in~\cite{Meulen1977,Cover98,Kim12}. On the other hand, if the transmitter and/or receiver nodes are allowed to have more than a single antenna, there will be a Gaussian vector channel in which much higher spectral efficiency (through spatial multiplexing) and reliability (by multi-path diversity) can be achieved by exploiting the scattering medium between the transmitter and receiver antenna arrays~\cite{telatar95, foschini98}. 
Specifically for the MISO BCs, the achievable throughput was developed in~\cite{Caire03} based on Costa's writing on dirty paper approach~\cite{Costa83}  that achieves the sum-capacity for a two-user case with a single transmitting antenna. Then, the sum-capacity for a general multiuser MISO BC was established in~\cite{Goldsmith03, Tse03, Wei04, Weingarten06} by exploiting the uplink-downlink duality between the multiaccess channel (MAC) and the corresponding BC. 
By using more practical finite-alphabet constellations rather than Gaussian input signals, the transmission schemes that maximize the mutual information  between the BS and all the receivers of the multi-user BC are considered in~\cite{Rajan09,Knopp10, Xiao11, Gao12}.


The aforementioned information-theoretic analyses serve as a guideline for a general system design.
For a transmitter design, non-linear precoding techniques such as the dirty paper coding~(DPC) method can be used to approach the sum rate of the MISO BC~\cite{Tse03 , Wei04}. It was shown in~\cite{Caire03} that a successive interference cancellation procedure, namely the ZF-DPC, can be performed at the transmitter side to completely remove mutual interference between receivers.  Given the complexity of DPC,  the Tomlinson-Harashima precoding~(THP) method~\cite{Tomlinson71, Windpassinger04, Garcia14,Viswanathan03} serves as a suboptimal but practical approximation of DPC by introducing a modulo operation to transmitted symbols.  Despite the fact that there is a modulo loss~\cite{Cioffi05}, the transmitted symbols are guaranteed to have a finite dynamic range. 
All these precoding methods  were primarily devoted to improving the sum rate of multiuser MISO BCs. On the other hand, practical transmitter designs may also aim to improve signal quality at the receiver side.  Among such transmitter designs, linear precoding techniques receive tremendous attention because of their potential and simplicity. Using signal-to-noise ratio (SNR) as a design criterion, it was shown that transmitter beamforming can increase the received SNR in the multiuser MISO channel by performing optimization on the beamformer design and the power allocation scheme~\cite{Liu98, Schubert02, Schubert04}. In addition, by employing MMSE as a performance measure, an optimal precoder was proposed in~\cite{Hochwald05} with regularized channel inversion, which outperforms the ZF scheme when the channel condition number is large. By maximizing SLNR for all users simultaneously, a closed-form beamforming method was given in~\cite{Sayed07}. Even so, however, it was demonstrated that the ZF beamforming technique, which is simple to implement, can achieve most of the capacity in moderdate and high SNR regimes~\cite{Shamai08, Spencer04}. Comprehensive comparisons of these schemes can be found in~\cite{Samardzija03}. 

As we can see, interference has long been the central focus for meeting the ever increasing requirements on quality of service in modern and future wireless communication systems. The key to the understanding of multiuser communications is the understanding of interference. Traditional approaches, which treat interference as a detrimental phenomenon are, therefore, to suppress or eliminate it.  
The classical information-theoretic study on the two-user Gaussian interference channel~\cite{Kobayashi81} suggests us that we should treat the interference as noise when it is weak and that the optimal strategy is to decode the interference when it is very strong. In addition, when the level of interference is of the same order of the power of a desired signal, one good strategy is to suppress all the undesired interferences into a smaller space that has no overlap with the signal space~\cite{Cadambe08, Bresler10, Yihongwu15}. However, some recent innovative approaches, which consider interference as a useful resource, are, thus, to make use of it for developing energy efficient and secure 5G communication systems~\cite{YiLiu15, Ganzheng14mag, Masouros13}. 
For example, interference can be used for boosting up the desired signal~\cite{Masouros09, Masouros15, Masouros12feb}
for energy harvesting~\cite{ Ruizhang15, Zhuhan15, Krikidis14, Schober13,Ruizhang13, Kaibinhuang14, Zhiguoding14, Hossain14}
or to deteriorate the  signal of the eavesdropper  for secure communication~\cite{Ganzheng13}. 

Inspired by~\cite{Bresler10}, in this paper we consider the management of interference for BC by carefully designing communication signals. To better elaborate on our idea, we would like to revisit some early seminal work~\cite{Shu76, Shu78} of how to strategically take advantage of the finite alphabet properties of digital communication signals for managing interference for a two-user access binary channel. Essentially, the Kasami and Lin's main idea is to carefully design such two finite length codes for the two users that when any sum binary signal of the two user codewords is received in a noiseless environment, each individual user codeword can be uniquely decoded, as well as in a noisy case, the resulting error is able to be correctable. Specifically, such uniquely decodable code (UDC) was explicitly constructed  for a two-user binary ensure channel~\cite{Tilborg85, Ahlswede99}. Then, this important concept was extended to the design of UDC based on trellis modulation for an $N$-user binary multi-access channel~\cite{Chevillat81}, which allows a number of users to access a common receiver simultaneously and outperforms the time sharing method in terms of the probability of error. Furthermore, the design of trellis-coded UDC  was investigated in a complex number domain to extract the desired signal from the superposition of the signal and cochannel interferences~\cite{Yoshida97, Yoshida98, Yongacoglu04}. In addition, the concept of UDC was also exploited to design variety of multi-resolution modulation schemes for BC and it was shown that they not only outperform the frequency division scheme by properly designing the resulting constellation\cite{Vetterli93, leefangwei93, Bhargava07, malladi2012}, but also reduce the transmission delay of the network at the cost of increased transmitting power for fading channels~\cite{Bhargava07}. Recently, a pair of uniquely decodable constellations was designed to study the capacity region of a two user Gaussian multi-access channel~\cite{Harshan11}. 

Indeed, it is the above aforementioned factors that greatly motivate and enlighten us to look into interference from the perspective of signal processing. In this paper, we are interested in exploring a novel signal processing technique to manage interference for BC, which allows strongly interfered user signals to cooperate with each other as a common desired sum signal from which each individual user signal is able to be uniquely and efficiently decoded. Specifically, our main contributions of this paper can be summarized as follows:
\begin{enumerate}
 \item   An explicit  construction of UDCG, which can be considered as a UDC in the complex domain for a multiuser case, for general PAM and rectangular QAM constellations for \emph{any} number of users is proposed. 
The main difference between our UDCG design and all currently available UFC designs in literature is that in our UDCG design, the sum-constellation and all the user constellations are PAM and QAM constellations with different scales. It is because of this nice geometric structure that once the sum signal is received, each individual user signal can be efficiently and uniquely decoded (see Algorithms~1 and~2). 

\item Using the newly developed UDCG, we propose a novel NOMA transmission scheme called QAM-modulation division for the  multiuser MISO BC. 
First, an optimal beamformer that maximizes the minimum SNR between the two receivers is obtained in a closed-form for a two-user case. Then, for a general network topology with more than two receivers, a grouping-based transmitter design problem is also investigated, with ZF eliminating the inter-group interference, where the grouping policy, the beamformer design and power allocation strategy are addressed.  
 It is demonstrated that when the Hermitian angle of the two channel vectors is small, our proposed division method has a much lower probability of error, which confirms that the NOMA method with proper interference control is a promising technology for 5G communications.
\end{enumerate}
Our work can be considered as a concrete, simple and systematic design of the constellation for NOMA~\cite{Saito14, Zhiguoding15}, serving different users with different power levels, and it has a considerable spectral gain over the traditional methods.

 {\bf{Notations}}: 
 Matrices and column vectors are denoted by boldface characters with uppercase  (e.g., ${\mathbf A}$) and lowercase (e.g., ${\mathbf b}$), respectively. 
The transpose of ${\mathbf A}$ is denoted by ${\mathbf A}^T$ while the Hermitian transpose of ${\mathbf A}$ (i.e., the conjugate and transpose of ${\mathbf A}$) is denoted by
${\mathbf A}^H$. $\left\|\mathbf b\right\|$ denotes the Euclidean norm of $\mathbf b$.
We let $j=\sqrt{-1}$ and $\binom{n}{k}=\frac{n!}{k!(n-k)!}$ be the binomial coefficient.
In addition, $\prod$ stands for the production operation and $\mathcal{A} \setminus \mathcal{B}=\{x \in \mathcal{A} ~{\rm and}~ x \notin \mathcal{B}\}$.
$\lfloor a \rfloor$ is the floor function which represents the largest integer no more than $a$ and $\lceil b \rceil$ is the ceiling function that returns the smallest integer not less than $b$. $a\mod b = a - b\times{\rm fix}(a/b)$ where ${\rm fix}(\cdot)$ rounds the variable to the nearest integer towards zero. 

\section{Uniquely Decomposable Constellation Group}
In this section, we introduce the definition of the UDCG and then, provide the flexible and the explicit construction of the UDCG using commonly-used PAM and QAM constellations and the corresponding efficient decoding algorithms.
\begin{definition}[UDCG]\label{def:audcg}
A group of constellations $\{\mathcal{X}_i\}_{i=1}^N$ is said to form a UDCG, denoted by $\{ \sum_{i=1}^N x_i: x_i \in \mathcal{X}_i\}=\uplus_{i=1}^N \mathcal{X}_i = \mathcal{X}_1 \uplus \mathcal{X}_2 \uplus \ldots \uplus \mathcal{X}_N$ , if there exist two groups of $x_i, \tilde x_i \in \mathcal{X}_i$ for $i=1, 2, \cdots, N$ such that $\sum_{i=1}^N x_i =\sum_{i=1}^N \tilde x_i$, then, we have $x_i =\tilde x_i$ for $i=1, 2, \cdots, N$.~\hfill\QED
\end{definition}
For presentation convenience, constellation $\uplus_{i=1}^N \mathcal{X}_i$ in Definition~\ref{def:audcg} is called the \emph{sum-constellation} of all $\mathcal{X}_i$ and each $\mathcal{X}_i$ is called the \emph{$i$-th sub-constellation} of $\uplus_{i=1}^N \mathcal{X}_i$ or \emph{$i$-th user constellation}. The concept of UDCG can be considered as an extension of UDC in binary field to the complex number domain for $N$-users~\cite{Shu76, Shu78, Tilborg85, Ahlswede99, Chevillat81, Yoshida97, Yoshida98, Yongacoglu04, Vetterli93, leefangwei93, Bhargava07, malladi2012, Harshan11}. However, we would like to emphasize  a major difference between the definition of UDCG and the traditional concept of UDC. In our Definition~1, we are interested in each analysis component of the decomposition, i.e., the geometrical structure of each user-constellation,  as well as in the synthesis component of the decomposition, i.e.,  the geometrical structure of the sum constellation.

The following property reveals such a fact that checking whether or not a group of constellations forms a UDCG is equivalent to checking whether or not the cardinality of the sum constellation is equal to the product of the cardinalities of the user-constellations. 
\begin{property}[Unambiguity]\label{unambiguity}
Given a group of constellations $\{\mathcal{X}_i\}_{i=1}^N$ with each having a finite size, if we let $\mathcal{G} =\{ \sum_{i=1}^N x_i : x_i \in \mathcal{X}_i \}$, then, $\mathcal{G} =\uplus_{i=1}^N \mathcal{X}_i$ if and only if $|\mathcal{G}| =\prod_{i=1}^N |\mathcal{X}_i|$.\hfill\QED
\end{property}
\emph{Proof}:  Let ${\mathcal Y}$ denote a set of $N$-tuples $\mathcal{Y}=\{(x_1,x_2,\ldots, x_N): x_i \in \mathcal{X}_i\}$. Then, $|\mathcal{Y}|=\prod_{i=1}^N |\mathcal{X}_i|$ by the combinatorial rule of product and $\mathcal{Y}$ is a finite set, since each $\mathcal{X}_i$ is a finite set.

If $|\mathcal{G}| =\prod_{i=1}^N |\mathcal{X}_i|$, then, we have $|\mathcal{G}| = |\mathcal{Y}|$. Since $\mathcal{G}$ and $\mathcal{Y}$ are finite sets, there exists a bijection map between these two sets, which is denoted by $f_{\rm bij}:\mathcal{G} \to \mathcal{Y}$~\cite{Ernest11}. 
Without loss of generality, we let $f_{\rm bij}(\sum_{i=1}^N x_i) =  (x_1,x_2,\ldots, x_N)$. As $f_{\rm bij}:\mathcal{G} \to \mathcal{Y}$ is a bijective map, then,  if $\sum_{i=1}^N x_i = \sum_{i=1}^N \tilde x_i$, we must have $(x_1,x_2,\ldots, x_N) =(\tilde x_1, \tilde x_2,\ldots, \tilde x_N)$ and hence, $x_i= \tilde x_i$ for $ i=1, 2, \cdots, N$. Then, by Definition~\ref{def:audcg}, we have $\mathcal{G} =\uplus_{i=1}^N \mathcal{X}_i$.

If $\mathcal{G} =\uplus_{i=1}^N \mathcal{X}_i$, by Definition~\ref{def:audcg}, for any $(x_1,x_2,\ldots, x_N), (\tilde x_1, \tilde x_2,\ldots, \tilde x_N)\in \mathcal{Y}$ satisfying $\sum_{i=1}^N x_i =\sum_{i=1}^N \tilde x_i$, we have $x_i =\tilde x_i$ for $i=1, 2, \cdots, N$, or equivalently $(x_1,x_2,\ldots, x_N) =(\tilde x_1, \tilde x_2,\ldots, \tilde x_N)$. Therefore, there exists an injective function $f_{\rm inj}: \mathcal{Y} \to \mathcal{G}$ such that $f_{\rm inj}\big( (x_1,x_2,\ldots, x_N)  \big) = \sum_{i=1}^N x_i$. Hence, $|G| \ge |\mathcal{Y}|=\prod_{i=1}^N |\mathcal{X}_i|$. From the construction of $\mathcal{G}$, we know that $|\mathcal{G}| \le \prod_{i=1}^N |\mathcal{X}_i|$.
As a result, $|\mathcal{G}| = \prod_{i=1}^N |\mathcal{X}_i|$. This completes the proof of the property.\hfill$\Box$  
Since PAM and QAM constellations are commonly used in modern digital communication systems, in this paper we are interested in uniquely decomposing them into the sum of a group of scaled PAM or QAM constellations.

\begin{theorem}[PAM]\label{AUDCGPAM}
Given two positive integers $K$ and $N$, let $K_i$ be any $N$ nonnegative integers satisfying $\sum_{i=1}^N K_i=K$.
Then, $2^K$-ary PAM constellation $\mathcal{G} = \{\pm (m-\frac{1}{2}) : m =1,2,\ldots, 2^{K -1}\}$ can be uniquely decomposed into the sum of $N$ sub-constellations ${\mathcal X}_i$ for $i=1, 2, \cdots, N$, i.e., $\mathcal{G} = \uplus_{i =1}^N \mathcal{X}_i$, where  
\begin{align*}
\mathcal{X}_1 = \begin{cases} \{0\} & K_1=0\\ \{\pm (m-\frac{1}{2}) : m =1,2,\ldots, 2^{K_1 -1}\} & K_1 \ge 1 \end{cases}
\end{align*} 
and for $i \ge 2$, 
\begin{align*}
\mathcal{X}_i = 
\begin{cases} \{0\} & K_i=0\\ \{\pm (m-\frac{1}{2}) \times 2^{\sum_{n=1}^{i-1} K_n }  \\ \qquad \qquad \qquad : m =1,2,\ldots, 2^{K_i -1}\}& K_i \ge 1
 \end{cases}.
\end{align*} 
\hfill\QED
\end{theorem}
\emph{Proof:} On one hand, we notice that $\sum_{i=1}^N x_i \in \mathcal{G}$, for any $x_i\in \mathcal{X}_i, \forall i\in\{1,2,\ldots, N\}$. On the other hand, 
we also note that $|\mathcal{X}_i| =\begin{cases}  1 &K_i=0\\ 2^{K_i} & K_i \ge 1 \end{cases}, \forall i\in\{1,2,\ldots, N\}$ and $|\mathcal{G}| =2^K$. Since $K =\sum_{i=1}^N K_i$, we have $|\mathcal{G}| = \prod_{i=1}^N |\mathcal{X}_i|$. By Property~\ref{unambiguity}, we attain $\mathcal{G} = \uplus_{i =1}^N \mathcal{X}_i$.
This completes the proof of Theorem~\ref{AUDCGPAM}.~\hfill$\Box$

Theorem~\ref{AUDCGPAM} reveals a significant property on the PAM constellation that any PAM constellation of large size can be uniquely decomposed into the sum of a group of the scaled version of the PAM constellations of variety of small sizes. Furthermore, the following algorithm proceeds to uncover an important advantage of such unique decomposition.         
\begin{algorithm_c}[Fast detection of PAM UDCGs]\label{fastpam}
Given a UDCG $\mathcal{G} = \uplus_{i =1}^N \mathcal{X}_i$ generated from a PAM constellation by Theorem~\ref{AUDCGPAM}.
For an observed noisy real signal $y=\sum_{i=1}^N x_i +\xi$, where $x_i \in \mathcal{X}_i$ and $\xi\sim \mathcal{N} (0, \sigma^2/2)$ is a real additive white Gaussian noise. Then, we have a fast detection method for estimating all user-signals stated as follows:
\begin{enumerate}
\item Quantization of the sum signal: Given $y$, the optimal estimate of  $g=\sum_{i=1}^N x_i$ is given as follows:
\begin{eqnarray}\label{eqn:sum_ml}
&&\hat{g}= \arg_{g \in \mathcal{G}} \min |y - g|\nonumber\\
&&=\left\{
\begin{array}{ll}
-\frac{2^K-1}{2},~~~~~~~~~~~~~~~~~~~~~y\le-\frac{2^K}{2};\\
\lfloor y+\frac{2^K}{2}\rfloor-\frac{2^K-1}{2},-\frac{2^K}{2}<y\le\frac{2^K}{2};\\
\frac{2^K-1}{2},~~~~~~~~~~~~~~~~~~~~~~~~~y>\frac{2^K}{2}.
\end{array}
\right.
\end{eqnarray}
\item Decoding of the user-signals: Let $\hat{g}$ be defined by~\eqref{eqn:sum_ml}. Then,
the estimates of all the original user-signals $\hat{x}_i$, satisfying $\sum_{i =1}^N \hat x_i=\hat g$, are uniquely determined as
\begin{align}\label{eqn:user1}
\!\! \!\! \hat x_1 =  \begin{cases} 0 & \! \! K_1 =0 \\
( \hat g +\frac{2^K-1}{2})  \!\! \! \! \mod 2^{K_1}  -\frac{2^{K_1}-1}{2} & \!\! K_1\ge 1
\end{cases}
\end{align}
and for $i \ge 2$,
\begin{align}\label{eqn:usern}
\!\! \!\!  \hat x_i =\begin{cases}
0 &K_i =0\\
  \Big(\frac{   \hat g +\frac{2^{K}-1}{2}  -  ( \hat g+\frac{2^K-1}{2}) \mod 2^{\sum_{\ell=1}^{i-1} K_\ell } }{2^{\sum_{\ell =1}^{i-1}K_\ell}  } \\ \quad  \mod 2^{K_i} -\frac{2^{K_i}-1}{2} \Big) 2^{\sum_{\ell=1}^{i -1}K_\ell} & K_i \ge 1
\end{cases}.
\end{align}
\end{enumerate}
\hfill\QED
\end{algorithm_c}

The proof of Algorithm\,\ref{fastpam} is given in Appendix\ref{appendix:algorithm1}.
As we know, a rectangular QAM constellation is generated from a pair of the PAM constellations. Hence, Theorem~\ref{AUDCGPAM} can be extended to the PAM and QAM mixed case in a straightforward manner, whose proof, therefore, is omitted. 
\begin{theorem}[QAM]\label{AUDCGQAM}
For two positive integers $N$ and $K =K^{(c)}+K^{(s)}$, with $K^{(c)}$ and $K^{(s)}$ being nonnegative integers, let $K_i^{(c)}$ and $K_i^{(s)}$ for $i=1, 2, \cdots, N$ denote any two given nonnegative integer sequences satisfying $K^{(c)} =\sum_{i=1}^N K_i^{(c)}$ and $K^{(s)} =\sum_{i=1}^N K_i^{(s)}$ with $K_i^{(c)}+K_i^{(s)}>0$. Then, there exists a PAM and QAM mixed constellation ${\mathcal Q}$ such that $\mathcal{Q} = \uplus_{i =1}^N \mathcal{X}_i$, where  $\mathcal{X}_i= \mathcal{X}^{(c)}_i \uplus j\mathcal{X}^{(s)}_i$, with $j\mathcal{X}^{(s)}_i =\{jx:x\in\mathcal{X}^{(s)}_i\}$. In addition, $\mathcal{Q}^{(c)} =\uplus_{i =1}^N \mathcal{X}_i^{(c)}$ and $\mathcal{Q}^{(s)} =\uplus_{i =1}^N \mathcal{X}_i^{(s)}$ are two PAM UDCGs given in Theorem~\ref{AUDCGPAM}  according to the rate-allocation $K_i^{(c)}$ and $K_i^{(s)}$, respectively.~\hfill\QED
\end{theorem}

Similar to Algorithm~\ref{fastpam},  we also have an efficient detection method for a UDCG based on the QAM constellation below:
\begin{algorithm_c}[Fast detection of QAM UDCG]\label{fastqam}
Let a UDCG $\mathcal{G} = \uplus_{i =1}^N \mathcal{X}_i$ be generated from the QAM constellation by Theorem~\ref{AUDCGQAM}.
Then, for an observed noisy complex signal $y=\sum_{i=1}^N x_i +\xi$, where $x_i \in \mathcal{X}_i$ and $\xi \sim \mathcal{CN} (0, \sigma^2)$ is an additive circularly-symmetric complex Gaussian noise, all the user-signals are efficiently estimated using the following two successive steps:
\begin{enumerate}
\item Quantization of the sum signal:  Let $y = y^{(c)} +j y^{(s)}$. Find the quantized PAM  signal $\hat g^{(c)} \in \mathcal{Q}^{(c)}$ and $\hat g^{(s)} \in \mathcal{Q}^{(s)}$ of $y^{(c)}$ and $y^{(s)}$, respectively by solving the following optimization problems
\begin{align*}
&\hat g^{(c)}= \arg_{g \in \mathcal{Q}^{(c)}} \min |y^{(c)} - g|, \\
&\hat g^{(s)} = \arg_{g \in \mathcal{Q}^{(s)}} \min |y^{(s)} - g|.
\end{align*}
\item Decoding of the user-signals: By Algorithm~\ref{fastpam}, the estimates of all the user real signals $\hat x_i^{(c)} \in \mathcal{X}_i^{(c)}$ and $\hat x_i^{(s)} \in \mathcal{X}_i^{(s)}$ for $i=1, 2, \cdots, N$ can be efficiently obtained such that $\hat g^{(c)}=\sum_{i=1}^N \hat x^{(c)}_i$ and $\hat g^{(s)} = \sum_{i=1}^N \hat x^{(s)}_i$ , and thus,  $\hat g_i = \hat x^{(c)}_i + j \hat x^{(s)}_i$.~\hfill\QED
\end{enumerate}
\end{algorithm_c}
\section{Modulation Division for Two-User MISO BC}
Our primary purpose in this section is to apply the UDCG based on the QAM constellation to the design of an optimal beamformer for a two-user BC. Toward this goal, let us specifically consider a MISO BC having two single-antenna receivers and a BS equipped with $M$ antennas which transmits independent and identically distributed (i.i.d.) signals $s_1$ and $s_2$ simultaneously to the two receivers. The channel is assumed to be flat fading and quasi-static. Let $\mathbf{h}_1=[h_{1,1}, h_{2,1}, \ldots, h_{M,1}]^H$ and $\mathbf{h}_2=[h_{1,2}, h_{2,2}, \ldots, h_{M,2}]^H$ denote, respectively, the channel links between BS and user 1 and 2,  which are perfectly available at the transmitter. 
Here, our main idea is that BS treats the two channels to be strongly interfered each other and hence, in order to serve the two receivers at the same time, the BS transmits a sum signal $s=s_1+ s_2$, with \emph{one} common beamforming vector $\mathbf{w} \in \mathbb{C}^{M\times1}$ to be designed, where $s_1$ and $s_2$ are randomly chosen from an aforementioned UDCG $\mathcal{Q} = \mathcal{X}_1 \uplus \mathcal{X}_2$ such that $s_1\in \mathcal{X}_1$ and  $s_2\in \mathcal{X}_2$. Then, the signal intended for each receiver can be decoded separately by using our fast detection method described in Algorithm~\ref{fastpam} or~\ref{fastqam}.
\subsection{Modulation Division for Two-User Case}
The equivalent complex-baseband channel model for the received signals at the two receivers is given by
\begin{align*}
y_1 = \mathbf{h}_1^H \mathbf{w}s  + \xi_1,\\
y_2 = \mathbf{h}_2^H \mathbf{w}s + \xi_2,
\end{align*}
where $s$ is the information carrying symbol for both users with $\mathbb E[|s|^2]=1$ and hence the total transmitted power is $P=\mathbb E[ |s|^2] \mathbf{w}^H \mathbf{w} \mathbb = \mathbf{w}^H \mathbf{w}$. Also, $\xi_1, \xi_2 \sim \mathcal{CN}(0, \sigma^2)$ are additive circularly-symmetric complex Gaussian noise arising at each receiver.  It is worth noting that the case where different receivers have different noise levels can be incorporated into our model by performing a scaling operation on the channel coefficients. Hence, the noises are assumed to be of equal variance.
The SNRs for the sum signal $s$ at each receiver are expressed by
 \begin{align*}
{\rm SNR}_{{\rm md}_1} =\frac{|\mathbf{h}_1^H \mathbf{w} |^2}{\sigma^2}, \qquad
{\rm SNR}_{{\rm md}_2} =\frac{|\mathbf{h}_2^H \mathbf{w} |^2}{\sigma^2}.
\end{align*}
%
By using a max-min fairness on the received SNR, we aim to solve the following optimization problem:
\begin{problem}\label{pbm:twousr}
Find the beamforming vector ${\mathbf w}$ such that
\begin{subequations}\label{optproblem2usr}
\begin{align}
&\max_{\mathbf{w}}\min~ \{\mathbf{w}^H \mathbf{h}_1 \mathbf{h}_1^H \mathbf{w} , \mathbf{w}^H \mathbf{h}_2 \mathbf{h}_2^H \mathbf{w}\},\\
&{\rm ~s.t.~~}\mathbf{w}^H\mathbf{w} =P.
\end{align}
\end{subequations}\hfill\QED
\end{problem}
Without loss of generality, we assume that $\|\mathbf{h}_1\|, \|\mathbf{h}_2\| \neq 0$, since otherwise, the solution is trivial and in fact  we can not achieve reliable communication to both users simultaneously in this case. 
Now, let
\begin{align}\label{eqn:amatrix}
\mathbf{A} =\mathbf{h}_1 \mathbf{h}_1^H- \mathbf{h}_2 \mathbf{h}_2^H.
\end{align}
\begin{enumerate}
\item If  $\mathbf{h}_1$ and $\mathbf{h}_2$ are linearly dependent (or equivalently, $\mathbf{h}_1 =\tau \mathbf{h}_2$ for some $\tau \in \mathbb C$), then $\mathbf{A}=(|\tau|^2 -1) \mathbf{h}_2 \mathbf{h}_2^H$. Hence, if $|\tau|=1$, we have $\mathbf{A}=\mathbf{0}$. Otherwise, $\mathbf{A}$ has rank one.
\item If $\mathbf{h}_1$ and $\mathbf{h}_2$ are linearly independent (i.e., $\mathbf{h}_1 \neq \tau \mathbf{h}_2, \forall \tau \in \mathbb C$), then, the rank of $\mathbf{A}$ is 2.
Let the eigenvalue decomposition of $\mathbf{A}$ be given by
\begin{align}\label{en:svdchn}
\mathbf{A} = \mathbf{V}\mathbf{\Sigma}\mathbf{V}^H,
\end{align}
where $\mathbf{V}$ is a unitary matrix and $\mathbf{\Sigma} ={\rm diag}(\lambda_1, -\lambda_2, 0, \ldots,0)$ with $\lambda_1>0$ and $\lambda_2 >0$. 
From~\eqref{en:svdchn}, we can obtain
\begin{align}\label{convertedchannel}
\mathbf{\Sigma} &
=\mathbf{V}^H ( \mathbf{h}_1 \mathbf{h}_1^H- \mathbf{h}_2 \mathbf{h}_2^H)\mathbf{V}
=\tilde{\mathbf{h}}_1\tilde{\mathbf{h}}_1^H -\tilde{\mathbf{h}}_2\tilde{\mathbf{h}}_2^H,
\end{align}
where $\tilde{\mathbf{h}}_1 = \mathbf{V}^H \mathbf{h}_1 = [\tilde{h}_{1,1},\tilde{h}_{1,2},0,\ldots, 0]^T$ and $\tilde{\mathbf{h}}_2 = \mathbf{V}^H \mathbf{h}_2 = [\tilde{h}_{2,1},\tilde{h}_{2,2},0,\ldots, 0]^T$. We also denote $\tilde{\mathbf{w}} = \mathbf{V}^H \mathbf{w}=[\tilde w_1,\tilde w_2,\ldots, \tilde w_M]^T$ and
\begin{align}\label{eqn:eqvchnmtx}
 \tilde{\mathbf{H}} = \mathbf{V}^H \mathbf{H}=[\tilde{\mathbf{h}}_1~\tilde{\mathbf{h}}_2].
\end{align}
Equation~\eqref{convertedchannel} is equivalent to 
\begin{subequations}\label{eqchannelrelation}
\begin{align}
&| \tilde{h}_{1,1}|^2 - |\tilde{h}_{2,1}|^2=\lambda_1\label{eqchannel1},\\
&| \tilde{h}_{1,2}|^2 - |\tilde{h}_{2,2}|^2=-\lambda_2\label{eqchannel2},\\
&\tilde{h}_{1,1}\tilde{h}_{1,2}^*=\tilde{h}_{2,1}\tilde{h}_{2,2}^*.\label{eqchannel3}
\end{align}
\end{subequations}
The above relationships can be characterized by
\begin{align}\label{parameterization}
\!   \! \! \!\! \! \! \! \! \! \! \!\begin{bmatrix}\tilde{h}_{1,1} &  \! \! \!  \tilde{h}_{2,1}\\  \tilde{h}_{1,2} & \! \! \! \tilde{h}_{2,2}\end{bmatrix}
=\begin{bmatrix}\sqrt{\lambda_1}\sec\theta e^{j\beta} &\!\! \! \! \sqrt{\lambda_1}\tan \theta e^{j(\gamma+\alpha)}\\ \sqrt{\lambda_2}\tan\theta e^{j(\beta-\alpha)}& \! \!\!\!\sqrt{\lambda_2}\sec\theta e^{j\gamma}\end{bmatrix}
\end{align}
where $\theta =\arccos \frac{\sqrt{\lambda_1} }{|\tilde{h}_{11}|}, 0\le \theta<\pi/2$ and $\beta = \arg(\tilde{h}_{1,1})$, $\gamma =\arg(\tilde{h}_{2,2})$ and $\alpha=\arg( \tilde{h}_{2,1} ) -\arg(\tilde{h}_{2,2})$.
\end{enumerate}
Now, we are ready to state one of our main results in this paper, i.e., the optimal solution to two-user beamforming problem in~\eqref{optproblem2usr}.
\begin{theorem}[Optimal beamforming for two-user cases]\label{thm:opttwousr}
Let $f(\mathbf{w}) = \min \{\mathbf{w}^H \mathbf{h}_1 \mathbf{h}_1^H \mathbf{w} , \mathbf{w}^H \mathbf{h}_2 \mathbf{h}_2^H \mathbf{w}\}$. Then, the optimal solution~$\mathbf{w}^{\rm opt}$ to Problem~\ref{pbm:twousr} is determined as follows:

 \emph{Scenario 1: $\mathbf{h}_1 =\tau \mathbf{h}_2, \tau \in \mathbb C$}.
\begin{align*}
\max_{\mathbf{w}^H\mathbf{w} =P}~f(\mathbf{w}) =\min \{P \|\mathbf{h}_1\|^2, P \|\mathbf{h}_2\|^2 \},
\end{align*}
where $\mathbf{w}^{\rm opt} = \frac{\sqrt{P}\mathbf{h}_1}{\|\mathbf{h}_1\|} = \frac{\sqrt{P}\mathbf{h}_2}{\|\mathbf{h}_2\|}$.

\emph{Scenario 2: $\mathbf{h}_1 \neq \tau \mathbf{h}_2, \forall \tau \in \mathbb C$}. The solution is given below:
\begin{enumerate}
\item $\lambda_1 \le \lambda_2$. Then
\!\!\!\!\!\! \!\!\!  \begin{enumerate}
\item for $0\le \sin \theta \le \frac{\lambda_1}{\lambda_2}$, we have
\begin{align*}
\max_{\|\mathbf{w}\|^2 =P}~f(\mathbf{w})  
=\frac{P\lambda_1\lambda_2}{\lambda_1+\lambda_2} \frac{(1+\sin\theta)^2}{\cos^2 \theta},
\end{align*}
where $\mathbf{w}^{\rm opt}=\mathbf{V}\tilde {\mathbf w}^{\rm opt}$ with $ \tilde{\mathbf w}^{\rm opt} =\Big[ \sqrt{\frac{P\lambda_2}{\lambda_1+\lambda_2}}e^{j\beta}, \sqrt{\frac{P\lambda_1}{\lambda_1+\lambda_2}}e^{j(\beta -\alpha)}, 0, \ldots,0 \Big]^T$.
\item for $\frac{\lambda_1}{\lambda_2}< \sin \theta < 1$, we have
\begin{align*}
\max_{\|\mathbf{w}\|^2 =P}~f(\mathbf{w}) = \frac{P(\lambda_1+\lambda_2\sin^2 \theta)}{\cos^2 \theta},
\end{align*}
where $\mathbf{w}^{\rm opt}=\mathbf{V}\tilde {\mathbf w}^{\rm opt}$ with $ \tilde{\mathbf w}^{\rm opt} = \Big[ \sqrt{\frac{P\lambda_1}{\lambda_1+\lambda_2\sin^2 \theta }}e^{j\beta}, \sqrt{\frac{P\lambda_2\sin^2 \theta}{\lambda_1+\lambda_2\sin^2 \theta }} e^{j(\beta-\alpha)}, 0, \ldots,0\Big]^T$.
\end{enumerate}
\item $\lambda_1 > \lambda_2$. Then, 
\begin{enumerate}
\item for $0\le \sin \theta \le \frac{\lambda_2}{\lambda_1}$, we have
\begin{align*}
\max_{\|\mathbf{w}\|^2 =P}~f(\mathbf{w}) =\frac{P\lambda_1\lambda_2}{\lambda_1+\lambda_2}\frac{(1+\sin\theta)^2}{\cos^2 \theta},
\end{align*}
where $\mathbf{w}^{\rm opt}=\mathbf{V}\tilde {\mathbf w}^{\rm opt}$ with $ \tilde{\mathbf w}^{\rm opt} =\Big[ \sqrt{\frac{P\lambda_2}{\lambda_1+\lambda_2}}e^{j(\gamma +\alpha)}, \sqrt{\frac{P\lambda_1}{\lambda_1+\lambda_2}}e^{j\gamma}, 0, \ldots,0 \Big]^T$.
\item and for $\frac{\lambda_2}{\lambda_1}< \sin \theta < 1$, we have
\begin{align*}
\max_{\|\mathbf{w}\|^2 =P}~f(\mathbf{w}) =\frac{P(\lambda_1\sin^2\theta+\lambda_2)}{\cos^2 \theta},
\end{align*}
where $\mathbf{w}^{\rm opt}=\mathbf{V}\tilde {\mathbf w}^{\rm opt}$ with $ \tilde{\mathbf w} ^{\rm opt}=\big[\sqrt{\frac{P\lambda_1 }{\lambda_1 +\lambda_2 \csc^2\theta}}e^{j(\gamma +\alpha)} , \sqrt{\frac{P\lambda_2 }{\lambda_1\sin^2\theta +\lambda_2 }}e^{j\gamma}, 0, \ldots,0\big]^T$.
\end{enumerate}
\end{enumerate}

\hfill\QED
\end{theorem}
The proof of Theorem~\ref{thm:opttwousr} can be found in Appendix\ref{appendix:theorem3}.  We would like to make the following comments on Theorem~\ref{thm:opttwousr}:
\begin{enumerate}
\item The problem dealt with in Theorem~\ref{thm:opttwousr} is different from the physical-layer multicasting problem discussed in~\cite{Luo06june}, where a group of users are interested in a common message. However, in our model, the information symbols intended for separate users are different and form a UDCG. In addition, despite the fact that the optimization problem in ~\cite{Luo06june} is more general, its solution is numerical and not necessarily global. Our Theorem~\ref{thm:opttwousr} gives the global solution in the closed form for the two-user case.  
\item  Here, it is should mentioned clearly that when we have finished our manuscript, we realize work~\cite{Choi15} dealing with the same optimization problem as ours for the optimal design of a multicast beamformer with superposition coding. Unfortunately, the optimal solution given in ~\cite{Choi15} holds only when the condition $0\le \sin \theta \le \frac{\min\{ \lambda_1, \lambda_2\}}{\max\{ \lambda_1, \lambda_2\}}$ in Theorem~\ref{thm:opttwousr} is satisfied. In other words, under that condition, the optimal solution is achieved in boundary ${\mathbf w}^H {\mathbf h}_1 {\mathbf h}_1{\mathbf w}={\mathbf w}^H {\mathbf h}_2 {\mathbf h}_2{\mathbf w}$, which, however, is not true in general. In fact, the condition under which the solution to the maximin optimization problem is reached in the boundary is widely and deeply studied in~\cite{Liang07oct}.
\end{enumerate}

\subsection{The Comparison between MD and ZF Method}
In this section, we compare the error performance of our proposed MD beamforming with that of ZF beamforming~\cite{Shamai08}. For simplicity, we assume that the information rates of the two receivers are the same
and that the channel matrix $\mathbf{H} =[\mathbf{h}_1 ~\mathbf{h}_2]$ has full column rank, whose singular values are $\sqrt{\mu_1}$ and $\sqrt{\mu_2}$  
with $\mu_1,\mu_2 > 0$.
Then, the received SNR for ZF beamforming with the max-min fairness criterion is determined by
\begin{align}\label{eqn:snrzf}
{\rm SNR}_{\rm zf} &= \frac{P}{\sigma^2\sum_{i=1}^2  [ (\mathbf{H}^H\mathbf{H})^{-1}]_{i,i}}=\frac{P\mu_1\mu_2}{\sigma^2(\mu_1+\mu_2)}.
\end{align}
On the other hand, for the MD method, by Theorem~\ref{thm:opttwousr}, the minimum received SNR between the two users for the sum signal is given by
\begin{align}
{\rm SNR}_{\rm md}=\frac{\max_{\|\mathbf{w}\|^2 =P}~f(\mathbf{w})}{\sigma^2}.
\end{align}
Jointly considering~\eqref{eqn:eqvchnmtx} and~\eqref{parameterization}, we can obtain
\begin{subequations}\label{eigrelation}
\begin{align}
\mu_1+\mu_2 &
                        ={\rm tr}(\tilde{\mathbf{H}}^H\tilde{\mathbf{H}})
                       =(\lambda_1+\lambda_2)\frac{1+\sin^2\theta}{\cos^2 \theta},\\
\mu_1\mu_2 &
                       =  \det (\tilde{\mathbf{H}}^H\tilde{\mathbf{H}})
                       =\lambda_1\lambda_2.                
\end{align}
\end{subequations}
Hence, \eqref{eqn:snrzf} can be further represented in terms of $\lambda_1$ and $\lambda_2$ as
\begin{align}
{\rm SNR}_{\rm zf} &= \frac{P\lambda_1\lambda_2 }{\sigma^2(\lambda_1+\lambda_2)}\frac{\cos^2 \theta}{1+\sin^2\theta}.
\end{align}
For discussion convenience, we define $ \kappa=\lambda_1/\lambda_2$
and the SNR gain as $\eta(\kappa, \theta) =10\log_{10}\frac{{\rm SNR}_{\rm md} }{{\rm SNR}_{\rm zf} }$.
Then, by Theorem~\ref{thm:opttwousr}, the SNR gain as a function of $\kappa$ and $\theta$ is given by
\begin{corollary}[SNR Gain in terms of $\kappa, \theta$]\label{cor:snrgain} The following statements are true: 
\begin{enumerate}
\item If $0<\kappa\le 1$ and $0\le \sin \theta \le \kappa$, then
$\eta(\kappa, \theta) 
=10\log_{10}\frac{1+\sin^2\theta}{(1-\sin\theta)^2}$;
\item If $0<\kappa\le 1$ and $\kappa< \sin \theta < 1$, then
$\eta(\kappa, \theta) 
=10\log_{10}\frac{(1+\sin^2\theta)(\kappa+\sin^2\theta)(1+1/\kappa)}{\cos^4\theta}$;
\item If $1<\kappa$, $0\le \sin \theta \le 1/\kappa$, then
$\eta(\kappa, \theta) 
=10\log_{10}\frac{1+\sin^2\theta}{(1-\sin\theta)^2}$;
\item If $1<\kappa$ and $1/\kappa< \sin \theta < 1$, then
$\eta(\kappa, \theta) 
=10\log_{10}\frac{(1+\sin^2\theta)(1/\kappa+\sin^2\theta)(1+\kappa)}{\cos^4\theta}$.
\end{enumerate}\hfill\QED
\end{corollary}
%
By Corollary~\ref{cor:snrgain}, the SNR gain can be evaluated once $\mathbf{H}$ has been obtained.
To further appreciate the physical meaning of the SNR gain, we have the following lemma.
\begin{lemma}\label{lemma:eigchn}
Given channel $\mathbf{H} =[\mathbf{h}_1 ~\mathbf{h}_2]$, let $\sqrt{\mu_1}$ and $\sqrt{\mu_2}$ denote its two singular values, $\|\mathbf{h}_1\|^2 =a$ and $\|\mathbf{h}_2\|^2 =b$, and $|\mathbf{h}_1^H \mathbf{h}_2|=c$. Also we let $\lambda_1$ and $\lambda_2$ be defined in~\eqref{en:svdchn}. Then, we have 
$\mu_1 = \frac{a+b +\sqrt{(a-b)^2 +4c^2}  }{2}$ ,
$\mu_2 = \frac{a+b -\sqrt{(a-b)^2 +4c^2}  }{2}$ and $\lambda_1 = \frac{a-b +\sqrt{(a+b)^2 - 4c^2}  }{2}$,
$\lambda_2 = \frac{-a+b+ \sqrt{(a+b)^2 - 4c^2}  }{2}$.
\hfill\QED
\end{lemma}
The proof of Lemma~\ref{lemma:eigchn} can be found in Appendix-\ref{appendix:lemma1}. By Lemma~\ref{lemma:eigchn}, we can immediately have the following corollary:
\begin{corollary}\label{cor:theta} Let $\theta$ be defined in~\eqref{parameterization}, and $a, b$ and $c$ be defined in Lemma~\ref{lemma:eigchn}. Then, 
we have $\sin \theta 
                =\frac{ 2c}{a+b+\sqrt{(a+b)^2 -4c^2}}$. 
\end{corollary}
 \hfill\QED
 
To gain more physical meaning of the SNR gain, we now define $\rho=a/b$ and $\cos \varphi =\frac{|\mathbf{h}_1^H \mathbf{h}_2|}{\|\mathbf{h}_1\| \|\mathbf{h}_2\|}$,
where $\varphi \in[0, \pi/2]$ is called the Hermitian angle~\cite{Scharnhorst01} between two channel vectors $\mathbf{h}_1$ and $\mathbf{h}_2$. The SNR gain as a function of $\rho$ and $\varphi$ is defined by
$\nu (\rho, \varphi) =10 \log_{10}\frac{{\rm SNR}_{\rm md} }{{\rm SNR}_{\rm zf} }$. Inserting $\theta$ in~ Corollary~\ref{cor:theta} into Corollary~\ref{cor:snrgain} and using Lemma~\ref{lemma:eigchn} and Corollary~\ref{cor:theta}, we can have the following corollary, whose proof is omitted. 
\begin{corollary}[SNR Gain in terms of $\rho, \varphi$]\label{corsnrgainrho} The following statements are true.
\begin{enumerate}
\item If $0<\rho \le 1$ and $0\le \cos\varphi\le\sqrt{\rho}$ (i.e., $0<\kappa\le 1$, $0\le \sin \theta \le \kappa$), then
$\nu (\rho, \varphi) 
=10\log_{10}  \frac{1+\rho}{1+ \rho-2\sqrt{\rho}\cos\varphi}$;
\item If $0<\rho \le 1$ and $\sqrt{\rho}< \cos \varphi \le 1$ (i.e., $0<\kappa\le 1$, $\kappa< \sin \theta < 1$), then
$\nu (\rho, \varphi) 
=10\log_{10} \frac{1+\rho}{1-\cos^2\varphi}$;
\item If $1<\rho$ and $0\le \cos\varphi \le 1/\sqrt{\rho}$ (i.e., $1<\kappa$, $0\le \sin \theta \le 1/\kappa$), then
$\nu (\rho, \varphi) 
=10\log_{10}  \frac{1+\rho}{1+ \rho-2\sqrt{\rho}\cos\varphi}$;
\item If $1<\rho$ and $1/\sqrt{\rho}<\cos\varphi \le 1$ (i.e., $1<\kappa$, $1/\kappa< \sin \theta < 1$), 
$\nu (\rho, \varphi) 
=10\log_{10} \frac{1+1/\rho }{1-\cos^2\varphi}$.
\end{enumerate}\hfill\QED
\end{corollary}

\begin{figure}[!hbp]
    \centering
    \flushleft
    \resizebox{8cm}{!}{\includegraphics{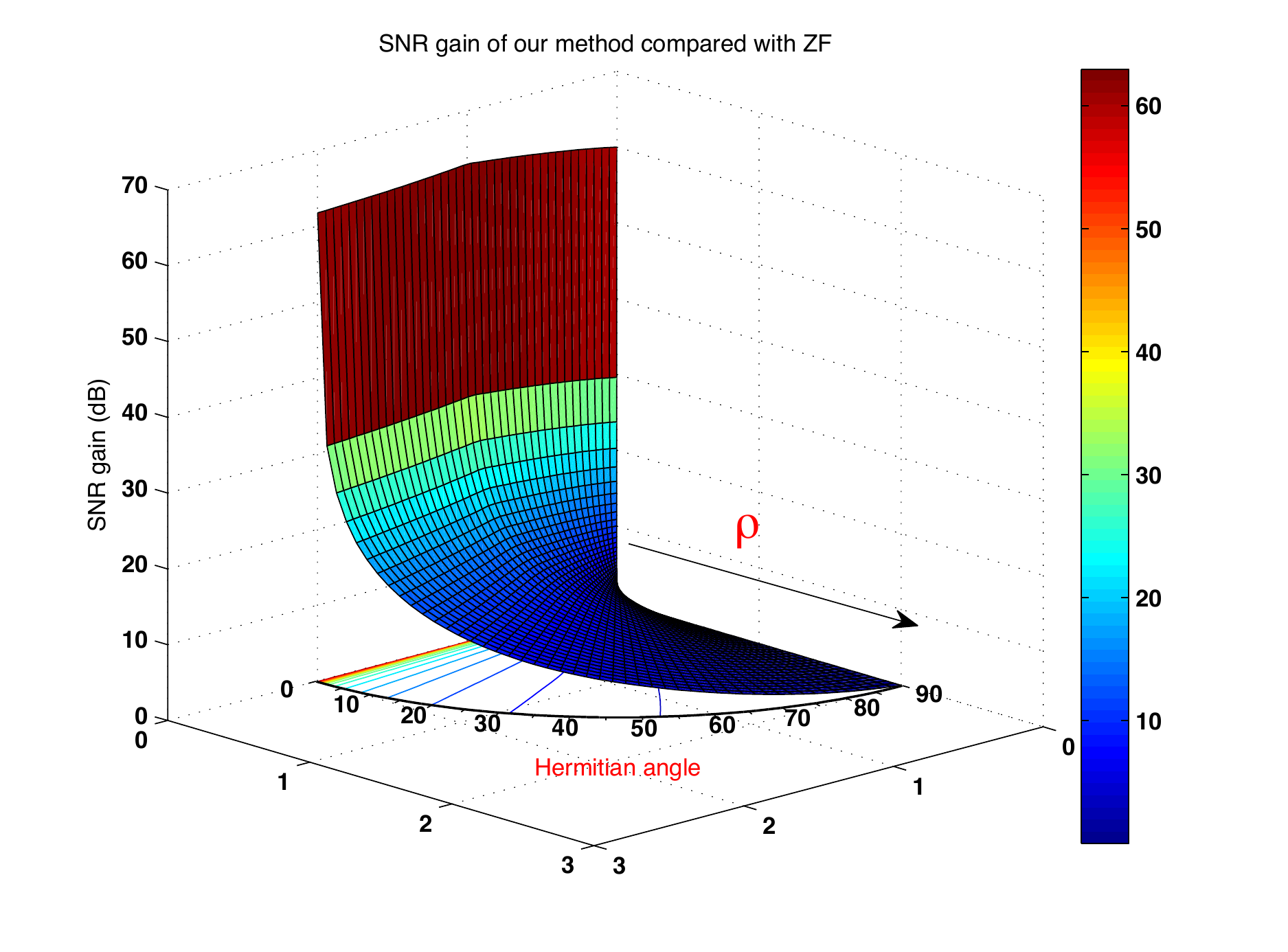}}
    \centering
    \caption{SNR Gain in terms of $\rho, \varphi$ in dB} 
    \label{fig:snrgain}
\end{figure}
From~Corrollary~\ref{corsnrgainrho}, it is not hard to obtain $\nu(\rho, \varphi) \ge 0$ for all $\rho>0, 0 \le \varphi \le \pi/2$. Hence, The SNR gain of our proposed MD beamforming is at least as good as that of ZF beamforming. To see it more clearly, 
the SNR gain in terms of $\rho$ and $\varphi$ are plotted in~Fig. \ref{fig:snrgain} for $0<\rho<2$ and $0<\theta<\frac{\pi}{2}$. It can be observed that for given $\rho$, the SNR gain is  determined by the Hermitian angle $\varphi$ between two channel vectors. When $\varphi$ approaches zero, i.e., $\mathbf{h}_1$ and $\mathbf{h}_2$ are approximately aligned with each other, the SNR gain is extremely large. For more clarity, the SNR gains for some specific cases are shown in Table~\ref{tab:snrgain} .
\begin{table}[!hbp]
\begin{center}
\begin{tabular}{|c|cccccc|}
\hline
$\varphi$ (rad)  & $ \frac{\pi}{180} $ & $ \frac{5 \pi}{180} $ & $ \frac{15 \pi}{180} $ & $ \frac{30 \pi}{180} $ & $ \frac{45 \pi}{180} $  & $ \frac{90\pi}{180} $ \\
\hline
$\rho=1/16$  & 35.43  & 21.46  & 12.00  &  6.28  &  3.27   &      0\\
\hline
$\rho=1/8$   & 35.67  &  21.71  &  12.25   &  6.53  &   3.52    &      0\\
\hline
$\rho=1/4$   & 36.13   & 22.16  &  12.71   &  6.99    & 3.98     &     0\\
\hline
$\rho=1/2$ & 36.92   & 22.96   & 13.50   &  7.78  &  4.77    &     0\\
\hline
$\rho=1$  & 38.17   & 24.20   & 14.68   &  8.73    & 5.33    & 0\\
\hline
\end{tabular}
\caption{SNR gain in term of $\rho$ and $\varphi$ in $\rm dB$}\label{tab:snrgain}
\end{center}
\end{table}

Corollary~\ref{corsnrgainrho} is very convenient for the SNR gain evaluation for the two-user case, since $\|\mathbf{h}_1\|^2, \|\mathbf{h}_2\|^2$, and $|\mathbf{h}_1^H \mathbf{h}_2|$ are very easy to compute.  As an example, we show how the SNR gain can be evaluated for LoS channels.
\begin{example}
The channel coefficients for a LoS channel~\cite{Tse05book} with two users are given by
\begin{subequations}
\begin{align*}
\mathbf{h}_1 &= \frac{\sqrt{a}e^{j\psi_1} }{\sqrt{M}}[1~e^{-j2\pi \Delta \Omega_1}~ e^{-j2\pi 2\Delta \Omega_1} \cdots ~e^{-j2\pi (M -1)\Delta \Omega_1}]^T,\\
\mathbf{h}_2 &=\frac{\sqrt{b}e^{j\psi_2}}{\sqrt{M}}[1~e^{-j2\pi \Delta \Omega_2}~ e^{-j2\pi 2\Delta \Omega_2} \cdots ~e^{-j2\pi (M -1)\Delta \Omega_2}]^T,
\end{align*}
\end{subequations}
where $a, b$ are the channel gain, $\Omega_1, \Omega_2$ are called the directional cosine with respect to the transmitting antenna array and $\Delta$ is the normalized transmitting antenna distance, normalized to the unit wavelength of carrier. %
Then, we have $\rho_{\rm LoS}=\frac{a}{b}$
and the Hermitian angle between two channel vectors $\varphi, \varphi \in (0,\pi/2)$ is determined by 
$\cos \varphi_{\rm LoS}  =\frac{|\mathbf{h}_1^H \mathbf{h}_2|}{\|\mathbf{h}_1\| \|\mathbf{h}_2\|}=\frac{1}{M}\Big|\frac{\sin(\pi M\Delta (\Omega_1-\Omega_2)) }{\sin( \pi \Delta(\Omega_1-\Omega_2) )}\Big|$.
By Corollary~\ref{corsnrgainrho}, the SNR gain can be computed against the directional cosine $\Omega_1, \Omega_2$ and the normalized antenna length $\Delta$. \hfill\QED
\end{example}

\section{Grouped Modulation Division Transmission for Multiuser MISO BC}
\begin{figure}[ht]
    \centering
    \flushleft
    \resizebox{8cm}{!}{\includegraphics{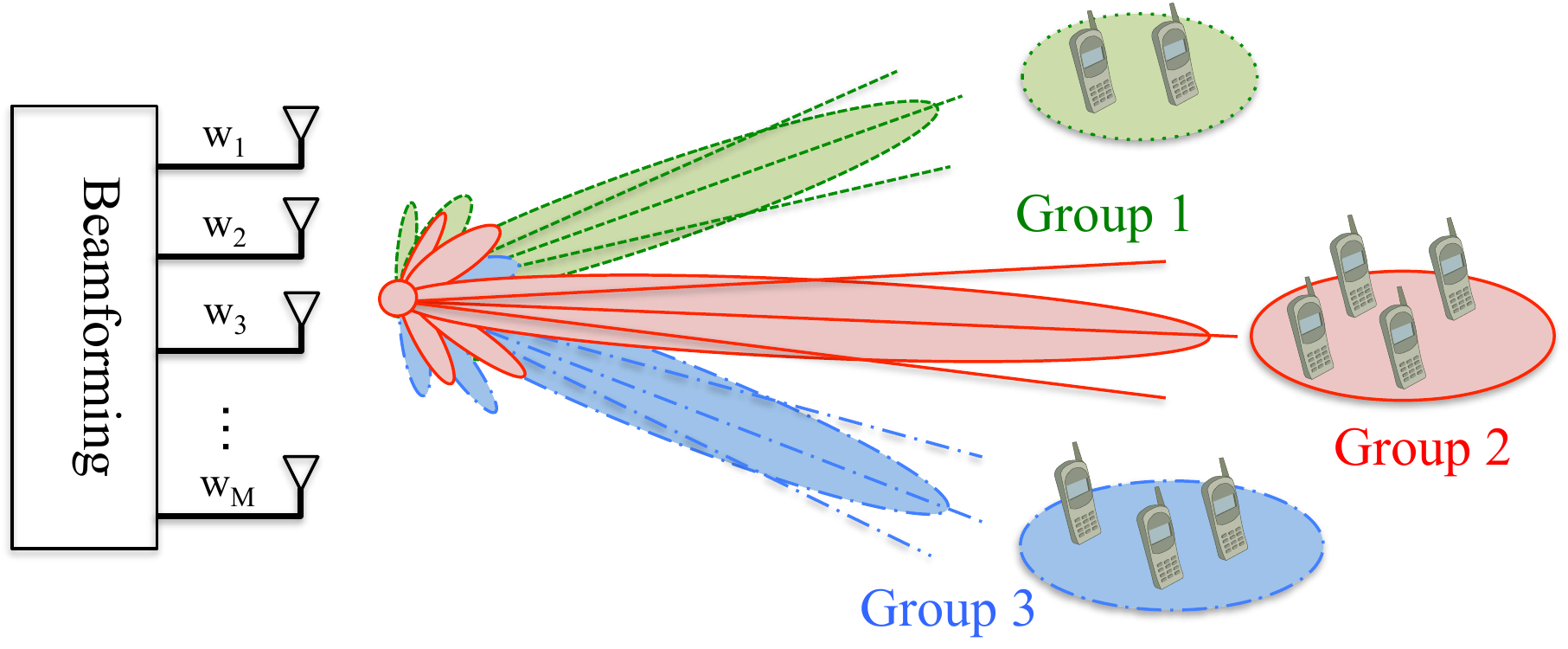}}
    \centering
    \caption{Illustration of Precoded MISO BC Model} 
    \label{fig:sysmodelside}
\end{figure}
In this section, a novel grouped modulation division transmission method is proposed for the multiuser MISO BC. The grouping algorithm is developed for cases where each group has at most two users. Then, the optimal beamforming vector and power allocation are all given in a closed-form.
\subsection{System Model}
We consider a communication system with a BS equipped with a set of $M$ transmitting antennas communicating with $N$ single antenna users $\mathcal{U} =\{U_1,U_2, \ldots, U_N\}$ simultaneously in the downlink as illustrated in Fig.~\ref{fig:sysmodelside}.  The channel links from BS to all the receivers can be stacked together into a matrix $\mathbf{H} \in \mathbb C^{M\times N}$, the $k$-th column of which is denoted by $\mathbf{h}_{k}=[h_{1,k}, h_{2,k}, \cdots, h_{M,k}]^H$, representing the channel link from BS to the $k$-th receiver.  The $N$ different receiver nodes can be further divided into $G\le N$ groups, with $k$-th group containing $N_k$ users, say, $U_{k_1}, U_{k_2}, \cdots, U_{k_{N_k}}$,  such that $N=\sum_{k=1}^G N_k$. For clarity, all the users are relabelled to represent the grouping results. If we let $\mathcal{S}$ denote a set consisting of all the users, then, it can be partitioned into $\mathcal{S} = \mathcal{S}_1 \cup \mathcal{S}_2\cup \ldots \cup \mathcal{S}_G$, $\mathcal{S}_k \cap \mathcal{S}_\ell= \varnothing, \forall k\neq \ell$, where $\mathcal{S}_k=\{U_{k_1},U_{k_2}, \ldots, U_{k_{N_k}}\}$. Correspondingly, the channel vector from BS to  $U_{k_\ell}$ is now denoted by $\mathbf{h}_{k_\ell}$ for $k \in\{1,2, \cdots, G\}$ and $\ell\in\{1, 2, \cdots, N_k\}$ and in turn, the channel matrix between BS and all the users in $\mathcal{S}_k$ is represented by
\begin{align}\label{eqn:chgroupk}
\mathbf{H}_k=[ \mathbf{h}_{k_1}, \mathbf{h}_{k_2}, \ldots, \mathbf{h}_{k_{N_k}}].
\end{align}
Meanwhile, the matrix containing channel links from BS to all the other users in $\mathcal{S}\setminus \mathcal{S}_k$ is represented by
\begin{align}\label{choutgroupk}
\bar{\mathbf{H}}_k=[\mathbf{H}_1,\ldots, \mathbf{H}_{k-1}, \mathbf{H}_{k+1},\ldots, \mathbf{H}_G ].
\end{align}
The grouping strategy of dividing the original user set $\mathcal{U}$ into $G$ mutually disjoint subsets $\mathcal{S}_k$ of $\mathcal{S}$ will be discussed later. For now, let us suppose that the grouping method has been given. Then, the communication process is carried out in the following two steps.  

Firstly, we assume that all the users in the group $\mathcal{S}_k$ use one UDCG $\mathcal{Q}_k = \uplus_{\ell =1}^{N_k} \mathcal{X}_{k_\ell}$, with each sub-constellation $\mathcal{X}_{k_\ell}$ adopted by user $U_{k_\ell}$. The rate allocation of group $\mathcal{S}_k$ is based on the sum decomposition $K_k =\sum_{\ell=1}^{N_k} K_{k_\ell}, \forall k\in\{1,2,\ldots, G\}$, where $K_k$ is the sum rate for all the users in group $\mathcal{S}_k$ and $K_{k_\ell} =\log_2 (|\mathcal{X}_{k_\ell} |)$ is the rate of user $U_{k_\ell}$. 

Secondly, a normalized information carrying signal $s_k$ intended for all the users in $\mathcal{S}_k$ is generated by using the UDCG $\mathcal{Q}_k = \uplus_{\ell =1}^{N_k} \mathcal{X}_{k_\ell}$, i.e.,
\begin{align*}
s_k = \frac{1}{ \sqrt{\mathbb{E} [|\sum_{\ell=1}^{N_k} s_{k_\ell}|^2]}} \sum_{\ell=1}^{N_k} s_{k_\ell},\quad \forall k\in\{1,2,\ldots, G\},
\end{align*}
where $s_{k_\ell}$ is assumed to be independently and uniformly drawn from the corresponding sub-constellation $\mathcal{X}_{k_\ell}$. It can be showed that the sum signal $s_k$ is also uniformly distributed over the scaled sum-constellation $\frac{1}{ \sqrt{\mathbb{E} [|\sum_{\ell=1}^{N_k} s_{k_\ell}|^2]}}\mathcal{Q}_k$ such that $\mathbb E[|s_k|^2]=1,\forall k$. It is worth pointing out that due to the power normalization, the minimum Euclidean distance of the constellation points of $s_k$ for different user group $\mathcal{S}_k$ might be different. Since the probability of error for the sum signal $s_k$ is dominated by the minimum Euclidean distance in high SNR regimes, it is anticipated that more transmitting power is required for $s_k$ when the sum-constellation is large with the same target error performance.

Then, all users in the same group $\mathcal{S}_k$ adopt the same precoding vector $\mathbf{w}_k$ and the weighted signals are fed into $M$ transmitter antennas at BS. Hence, the received signal at user $U_{k_\ell}$ can be expressed by
\begin{align}\label{eqn:receivedsignal}
y_{k_\ell} = \underbrace{ \mathbf{h}_{k_\ell}^H  \mathbf{w}_k s_k}_{{\rm intra-group~ signal}}\!\! +\underbrace{ \mathbf{h}_{k_\ell}^H \sum_{m=1, m\neq k}^{G} \mathbf{w}_m s_m }_{\rm out~of~group~interference} + \underbrace{\xi_{k_\ell}}_{\rm noise},
\end{align}
in which $\xi_{k_\ell}\sim \mathcal{CN}(0,\sigma^2)$ is the circularly-symmetric complex Gaussian noise arising at $U_{k_\ell}$.  Here, the noise variance is assumed to be the same for all the users. The case with different receiver noise level can be incorporated into our model by performing a scaling operation on the channel coefficient $\mathbf{h}_{k_\ell}$.

In the receiver side, all users $U_{k_\ell}$ can detect the information intended for themselves from the uniquely decomposable signal $s_k$ by using our fast detection method, i.e., Algorithms~\ref{fastpam} and~\ref{fastqam} while treating the out of group interference as additive noise. However, in high SNR regimes, the out of group interference is the dominant term that limites the error performance of our system. In what follows, a novel transmission scheme is proposed so that the out of group interference is completely cancelled out by using the ZF philosophy while the intra-group interference contained in~$s_k$ can be eliminated by taking advantage of the uniquely decomposable property of the sum-constellation.

\subsection{Weighted Max-Min Fairness Grouped Transmission with ZF and Modulation Division}
From~\eqref{eqn:receivedsignal}, we know that the cochannel interference for $U_{k_\ell}$ consists of two parts: 1) the inter-group interference (i.e., interference originated from users in $\mathcal{S} \setminus \mathcal{S}_k$) and 2) the intra-group interference (i.e.,  interference due to users in $ \mathcal{S}_k \setminus \{U_{k_\ell}\}$). In our scheme, we use a ZF  method to cancel the out of group interference, i.e.,
\begin{align}\label{zf:outofgroup}
\bar{\mathbf{H}}_k^H \mathbf{w}_k =\mathbf{0}, \forall k\in \{1,2, \ldots, G\},
\end{align}
where $\bar{\mathbf{H}}_k$ is defined in~\eqref{choutgroupk}.
Now, user $U_{k_\ell}$ only suffers from intra-group interference which can be also eliminated later by utilizing uniquely decomposable property. Under the ZF constraint~\eqref{zf:outofgroup}, the SNR for the sum signal $s_k$ at user $U_{k_\ell}$ can be expressed by ${\rm SNR}_{k_{\ell}} =\frac{ |\mathbf{h}_{k_\ell}^H  \mathbf{w}_k|^2 }{\sigma^2}, \forall k,\ell$.
Let us denote the total transmitted power for all the users in $\mathcal{S}_k$ at BS as
\begin{align}\label{pwgroupk}
P_k=  \mathbb E[| \mathbf{w}_k s_k|^2]=\mathbf{w}_k^H \mathbf{w}_k,\quad  \forall k\in \{1,2, \ldots, G\}.
\end{align}
Therefore, we aim to solve the following weighted max-min grouped beamforming optimization problem:
\begin{problem}\label{pbm:groupbf} Find the beamforming vectors ${\mathbf w}_k$ such that the worst case weighted received signal power is maximized, i.e.,
\begin{subequations}\label{eqn:opt1}
\begin{align}
&\max_{\mathbf{w}_k, \forall k}~\min_{\forall k,\ell} ~\varrho_{k} |\mathbf{h}_{k_\ell}^H  \mathbf{w}_k|^2 \\
&{\rm s.t.~} ~\eqref{zf:outofgroup}~{\rm and}~\sum_{k=1}^G \mathbf{w}_k^H \mathbf{w}_k= P.
\end{align}
\end{subequations}
~\hfill\QED
\end{problem}
The quantity $\varrho_{k} |\mathbf{h}_{k_\ell}^H  \mathbf{w}_k|^2$ in Problem~\ref{pbm:groupbf} is usually called the weighted received signal power for $s_k$ at $U_{k_\ell}$. Using weighted SNR is a common method to balance the QoS among different users~\cite{Luo08tsp, Ottersten14}.  The resulting ${\rm SNR}_{k_{\ell}} $ is anticipated to be increased with $\varrho_{k}$ being decreased. Since the error performance for each user is mainly determined by the SNR of the sum signal $s_k$ and the minimum Euclidean distance of the sum-constellation of user groups $\mathcal{S}_k$, in this paper we choose  $\varrho_{k}$ to be
\begin{align}\label{qosprevision}
\varrho_{k}=\frac{1}{\mathbb{E} [|\sum_{\ell=1}^{N_k} s_{k_\ell}|^2]}
\end{align}
which reasonably balances the minimum Euclidean distance for the sum signal $s_k$ in different user groups. Other choices of $\varrho_{k}$ are possible based on different application requirements. 

In order to solve Problem~\ref{pbm:groupbf}, we first examine whether or not its feasible domain, ${\mathcal W}=\{{\mathbf w}=({\mathbf w}_1^T, {\mathbf w}_2^T, \cdots, {\mathbf w}_G^T)^T:  \bar{\mathbf H}_k {\mathbf w}_k=0\,{\rm and}\, \sum_{k=1}^G {\mathbf w}_k^H {\mathbf w}_k=P\}$ is empty, i.e., Problem~\ref{pbm:groupbf} is feasible. This essentially checks whether constraint 
$
\bar{\mathbf{H}}_k^H \mathbf{w}_k =\mathbf{0}, k\in \{1,2, \ldots, G\},
$
can be satisfied. Since $\bar{\mathbf{H}}_k \in \mathbb C^{M \times (N-N_k)}$ has a rank of $N-N_k$, where $N_k\ge 1$,  the constraint can be satisfied if $M \ge N-N_k, \forall k$. This condition is indeed satisfied, since we assume $N\le M+1$ in this paper. Therefore, Problem~\ref{pbm:groupbf} is always feasible. On the other hand, we observe an important fact on the feasible domain. For any fixed $P_k$, $0\le P_k\le P$ for $k=1, 2, \cdots, G$, if we let ${\mathcal W}(P_1, P_2, \cdots, P_G)=\{({\mathbf w}_1^T, {\mathbf w}_2^T, \cdots, {\mathbf w}_G^T)^T:  \bar{\mathbf H}_k {\mathbf w}_k=\mathbf{0}\,{\rm and}\, {\mathbf w}_k^H {\mathbf w}_k=P_k,\,k=1,\, 2,\, \cdots,\, G \}$. Since ${\mathcal W}$ can be decomposed into a union of all such ${\mathcal W}(P_1, P_2, \cdots, P_G)$, i.e., ${\mathcal W}=\bigcup_{\sum_{k=1}P_k=P}{\mathcal W}(P_1, P_2, \cdots, P_G)$. Therefore,  the original optimization Problem~\ref{pbm:groupbf} can be equivalently split into the following two kinds of sub-optimization problems:

\underline{{\bf Sub-problem 2.1}}: For any fixed $P_k$, $0< P_k<P$, find the beamforming vectors ${\mathbf w}_k, \forall k\in \{1,2,\ldots G\}$ such that 
\begin{subequations}\label{eqn:opt2}
\begin{align}
\zeta(P_k)&= \max_{\mathbf{w}_k}\min_{\forall \ell}~ |\mathbf{h}_{k_\ell}^H  \mathbf{w}_k|^2\\
{\rm s.t.~} &\bar{\mathbf{H}}_k^H \mathbf{w}_k =\mathbf{0}~{\rm and}~\mathbf{w}_k^H\mathbf{w}_k =P_k, 
\end{align}
~\hfill\QED
\end{subequations}

\underline{{\bf Sub-problem 2.2}}: Once sub-problem 2.1 has been solved, find an optimal power allocation strategy for all user groups $\mathcal{S}$ such that
\begin{align}\label{eqn:opt3}
& \max_{P_k, \forall k}~\min_{\forall k}~\varrho_k \zeta(P_k) \qquad {\rm s.t.~} \sum_{k=1}^G P_k=P.
\end{align}
~\hfill\QED

In general, the optimization problem~\eqref{eqn:opt2} for arbitrary $N_k \ge 3, \forall k$ is hard to solve. However, since the power required for using a large sum-constellation $\mathcal{Q}_k = \uplus_{\ell =1}^{N_k} \mathcal{X}_{k_\ell}$ with certain error target is huge if $N_k$ is too large, in this paper we primarily restrict ourself in the case with $N_k\le 2$. In this case, we assume that $N\le M+1$. Let us consider~\eqref{eqn:opt2} first, where $P_k$ is temporarily regarded as a fixed number.  For group $\mathcal{S}_k$ with $N_k=1$,  by the Cauchy-Swarz inequality we have $\mathbf{w}_k =\sqrt{P_k}\frac{\mathbf{h}_{k_1}}{\|\mathbf{h}_{k_1}\|}$.
For user group $\mathcal{S}_k$ with $N_k=2$, the sub-optimization problem~2.1 can be reformulated as

\begin{align}\label{opt:bftwousr}
   \zeta(P_k)=&\max_{\|\mathbf{w}_k\|^2 =P_k}\min~\{ \mathbf{w}_k^H\mathbf{h}_{k_1}   \mathbf{h}_{k_1}^H  \mathbf{w}_k, \mathbf{w}_k^H\mathbf{h}_{k_2}   \mathbf{h}_{k_2}^H  \mathbf{w}_k   \}\nonumber \\
&{\rm s.t.~}  \bar{\mathbf{H}}_k^H \mathbf{w}_k =\mathbf{0}, \quad \forall k\in \{1,2,\ldots G\}.
\end{align}
Let us consider the constraint of~\eqref{opt:bftwousr} first.
 For $N_k=2$, we have $\bar{\mathbf{H}}_k \in \mathbb C^{M \times (N-2)}$, which is a tall matrix of full column rank, since $N\le M+1$. 
 This constraint essentially requires that $\mathbf{w}_k$ lies in the orthogonal complement subspace of ${\rm span}( \bar{\mathbf{H}}_k)$.
Since $\bar{\mathbf{H}}_k$ has full column rank, the orthogonal complement projector for ${\rm span}( \bar{\mathbf{H}}_k)$ is determined by $\mathbf{P}_k = \mathbf{I} - \bar{\mathbf{H}}_k (\bar{\mathbf{H}}_k^H \bar{\mathbf{H}}_k)^{-1}\bar{\mathbf{H}}_k^H \in \mathbb{C}^{M\times M}$,  where $\bar{\mathbf{H}}_k^H \mathbf{P}_k=\mathbf{0}$. We know that the rank of $\mathbf{P}_k$ is $(M-N+2)$. Now we want to find an orthonormal basis for $\mathbf{w}_k$.
To do that, let the QR-decomposition of $\mathbf{P}_k$ be $\mathbf{P}_k =\mathbf{Q}_k \mathbf{R}_k$, where $\mathbf{Q}_k\in \mathbb C^{M\times (M - N +2)}$ is a column-wise unitary matrix. If we let $\mathbf{w}_k =\mathbf{Q}_k \check{\mathbf{w}}_k$, then,  
problem~\eqref{opt:bftwousr} is equivalent to
\begin{align*}
  \zeta(P_k) =\max_{\|\check{\mathbf{w}}_k\|^2 =P_k}\min~\{ \check{\mathbf{w}}_k^H\check{\mathbf{h}}_{k_1} \check{\mathbf{h}}_{k_1}^H  \check{\mathbf{w}}_k,  \check{\mathbf{w}}_k^H\check{\mathbf{h}}_{k_2}  \check{\mathbf{h}}_{k_2}^H   \check{\mathbf{w}}_k   \} ,
\end{align*}
where $\check{\mathbf{h}}_{k_\ell} = \mathbf{Q}_k^H \mathbf{h}_{k_\ell}, \forall k,\ell$.
The above optimization problem can be solved by using Theorem~\ref{thm:opttwousr}, with the optimal value, i.e.,   $\zeta(P_k)$, being linear in terms of $P_k$, i.e., 
\begin{align}\label{zeta}
\zeta (P_k) =P_k \varsigma_k, 
\end{align}
in which $\varsigma_k $ is determined by channel coefficients and is independent of $P_k$.
Our next goal is to solve sub-problem 2.2 in this case. Substituting~\eqref{zeta} into~\eqref{eqn:opt3} yields 
\begin{align}
\max_{P_k, \forall k}~\min_{\forall k}~ P_k \varrho_k\varsigma_k \qquad
{\rm s.t.~} \sum_{k=1}^G P_k= P
\end{align}
Its optimal value is attained when all $P_k \varrho_k\varsigma_k$ for $k=1, 2, \cdots, G$ are equal to each other, hence, leading to
\begin{align*}
P_k^{\rm opt} =\frac{P \prod_{\ell\neq k}^{G} \varrho_\ell\varsigma_\ell }{\sum_{m=1}^{G}  \prod_{n \neq m}^{G} \varrho_n\varsigma_n}, \quad\forall k \in\{1,2, \ldots, G\}.
\end{align*}
Thus far, we have solved the problem~\eqref{eqn:opt1}  for given grouping method $\mathcal{S}$ with $N_k \le 2, \forall k\in\{1,2, \ldots, G\}$.
\subsection{User Grouping for $N_k \le 2, \forall k\in\{1,2, \ldots, G\}$}
As we have mentioned before, the performance of our proposed transmission method is closely related to the user grouping strategy.  In this subsection, we consider the user grouping method for cases with $N_k \le 2, \forall k\in\{1,2, \ldots, G\}$. In these cases, we require that $G\le N \le 2G$ and as a consequence, there are $N-G$ groups with each having 2 users and $2G-N$ groups with each having one user. For example, if $G=N$, each group has only one user. If $G=N/2$, each group has exactly two users. Since $\mathcal{S}$ and $\mathcal{S}_k$ are all unordered sets, for the given number of groups $G$, we have $\frac{\prod_{k=0}^{N-G-1} \binom{N-2k}{2}}{(N-G)!}$ different grouping methods if $G \le N-1$ and only one method if $G=N$. Since for $N_k \le 2$,  $\lceil N/2\rceil \le G\le N$, we have in total $1+\sum_{m= \lceil N/2\rceil}^{N-1} \frac{\prod_{k=0}^{N-m-1} \binom{N-2k}{2}}{(N-m)!}$ different grouping ways. For small $N$, the optimal grouping method can be found by brute-force search. However, it would be prohibitively complicated for large $N$, which makes our design hard to implement in practice. Therefore, we now propose a suboptimal user-grouping method to make trade-off between performance and complexity by setting a threshold $\gamma_{T}$, which is a predefined level to balance the error performance among different groups.
\begin{example}
Consider a two-user MISO BC with ZF beamforming, where each user employs a square $K$-ary QAM constellation and hence, the sum-rate of this network is $2\log_2 K$.
In contrast, for the modulation division method, the sum-constellation is set to be a $K^2$-ary QAM constellation such that the sum-rate is $\log_2 K^2 =2\log_2 K$, which is the same as ZF method. Assume that the average power of the transmitted symbol $x_k$ for the ZF method and that of $s_k$ for the sum-constellation are all unified to 1. Then the minimum Euclidean distance of the constellation points of $x_k$ is $d_{\rm zf}(K) = \sqrt{\frac{6}{K-1}}$ and that of $s_k$ is $d_{\rm md}(K^2) = \sqrt{\frac{6}{K^2-1}}$. As a consequence we can set $\gamma_{T}=10\log_{10} \frac{d^2_{\rm zf}(K) }{d^2_{\rm md}(K) } =10\log_{10} (K+1)$ to compensate the SNR loss due to using a larger constellation. 
For example, every user is using a 4-QAM, we would expect $\gamma_{T} =6.99$dB, $\gamma_{T} =12.30$dB for 16-QAM, $\gamma_{T} =18.13$dB for 64-QAM and $\gamma_{T} =24.10$dB for 256-QAM.
\hfill\QED
\end{example} 

\begin{algorithm_c}[Grouping method]\label{algorithm:suboptgroup}
The grouping method for $N_k \le 2, \forall k\in \{1,2, \ldots, G\}$ is given as follows for $N$ receivers such that $N=\sum_{k=1}^{N_k} N_k$.  There are $\binom{N}{2}=\frac{N(N-1)}{2}$ possible grouping methods for one group with exactly two users. 
\begin{enumerate} 
\item Enumeration: Find all the possible groups with two users and generate the grouping index.
\item Grouping gain calculation: For all the grouping indexes, calculate the coding gain. For example, suppose that users $m$ and $n$ ($m<n$) are grouped together with channel matrix $\check{\mathbf{H}}_{m,n}=[\mathbf{h}_m, \mathbf{h}_n]$ and another matrix containing all the channel links of users in $\mathcal{U} \setminus \{U_m,U_n\}$, which is denoted by $\breve{\mathbf{H}}_{m,n}=[\mathbf{h}_1,\ldots, \mathbf{h}_{m-1}, \mathbf{h}_{m+1}, \ldots, \mathbf{h}_{n-1}, \mathbf{h}_{n+1},\ldots ,\mathbf{h}_N]$. 
 Let $\breve{\mathbf{P}}_{m,n} = \mathbf{I} - \breve{\mathbf{H}}_{m,n} (\breve{\mathbf{H}}_{m,n}^H \breve{\mathbf{H}}_{m,n})^{-1}\breve{\mathbf{H}}_{m,n}^H$ and its QR-decomposition be $\breve{\mathbf{P}}_{m,n}=\breve{\mathbf{Q}}_{m,n} \breve{\mathbf{R}}_{m,n}$. Compute $\rho_{m,n}=\frac{\|\breve{\mathbf{h}}_m\|^2 }{\breve{\|\mathbf{h}}_n\|^2}$ and $\cos \varphi_{m,n} =\frac{|\breve{\mathbf{h}}_m ^H \breve{\mathbf{h}}_n|}{\|\breve{\mathbf{h}}_m \| \|\breve{\mathbf{h}}_n\|}$, where $\breve{\mathbf{h}}_m = \breve{\mathbf{Q}}_{m,n}^H \mathbf{h}_{m}$ and $\breve{\mathbf{h}}_n = \breve{\mathbf{Q}}_{m,n}^H \mathbf{h}_{n}$. Then, by Corollary~\ref{corsnrgainrho}, calculate the grouping gain $\nu_{m,n}=\nu(\rho_{m,n}$, $\cos \varphi_{m,n} )$ as a function of $\rho_{m,n}$, $\cos \varphi_{m,n}$ if users $m$ and $n$ are grouped together. 
\item Sorting:  Now the $\frac{N(N-1)}{2}$ grouping gains $\nu_{m,n}$ are sorted in descending order, forming a vector $[\nu_{m_1,n_1}, \nu_{m_2,n_2}, \ldots, \nu_{m_{\frac{N(N-1)}{2}}, n_{\frac{N(N-1)}{2}}}]^T$.  
\item Grouping:  If $\nu_{m_1,n_1}>\gamma_{T}$, then, users $m_1$ and $n_1$ are grouped together. Otherwise,  go to the next step and no users are grouped together~\footnote{Then our method essentially degrade into the ZF method.}. Then, consider $\nu_{m_2,n_2}$. Again, if $\nu_{m_2,n_2} \le \gamma_{T}$, go to the next step and terminate the grouping procedure.  Otherwise, if $\nu_{m_2,n_2} > \gamma_{T}$ and either $m_2$ or $n_2$ has not been grouped yet, then, $m_2$ and $n_2$ are grouped together.  Repeat this process until all the $\nu_{m,n}$ have already been considered. The remaining users is left ungrouped.

\item Stop and output the grouping index.
~\hfill\QED
\end{enumerate}
\end{algorithm_c}
In our model, the grouping operation is carried out at the BS and then, the grouping indexes are informed to all the receivers. In fact, each receiver $\mathcal{U}_{k_\ell}$ only  needs to know the grouping index $k_\ell$ to obtain the sum-constellation used by the group and the corresponding sub-constellation of itself.

\begin{figure}[ht]
    \centering
    \flushleft
    \resizebox{9cm}{!}{\includegraphics{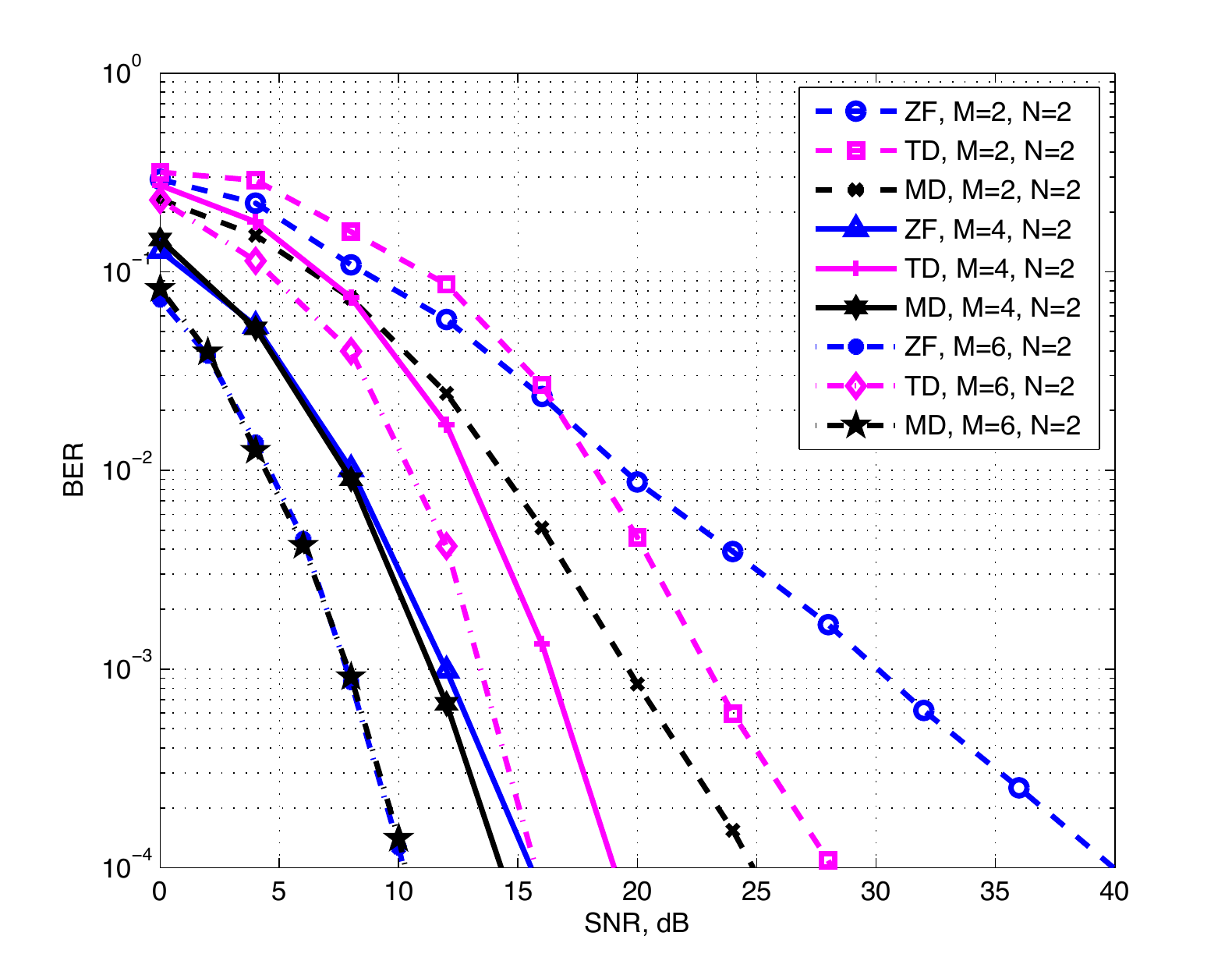}}
    \centering
    \caption{ BER performance against SNR with $M=2,4,6$, $N=2$ for i.i.d. Rayleigh channel, i.e., no transmitter correlation ($\rho=0$); Each user uses a 4-QAM.} 
    \label{fig:combined_n2_rayleigh}
\end{figure}
\begin{figure}
    \centering
    \flushleft
    \resizebox{9cm}{!}{\includegraphics{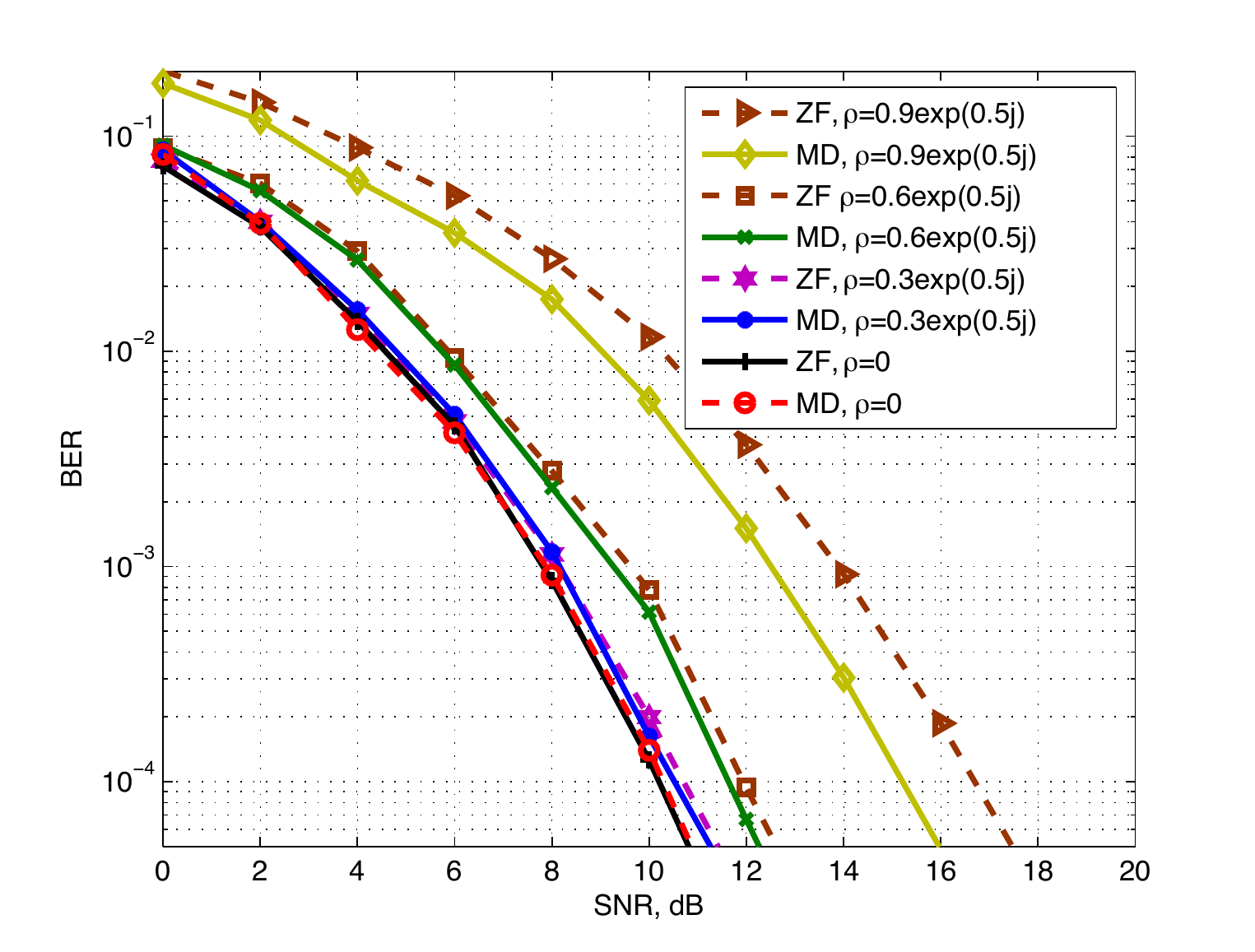}}
    \centering
    \caption{Average BER performance against SNR with $M=6$, $N=2$ with different $\rho$. Each user uses a 4-QAM.} 
    \label{fig:m6n2varrho}
\end{figure}
\section{Computer Simulations and Discussions}
In this section, computer simulations are carried out to verify our theoretical analysis and to assess the effectiveness of our proposed modulation division transmission method in a multiuser MISO BC. Throughout our simulations, we assume that the channel links from the BS are potentially correlated, but are uncorrelated between different users, i.e.,
\begin{align}
{\mathbb E}[ \mathbf{h}_k\mathbf{h}^H_\ell] =\begin{cases}
\mathbf{0} & \forall k \neq \ell,\\
\mathbf{\Sigma} & {\rm otherwise}
\end{cases}
\end{align}
\begin{figure}
    \centering
    \flushleft
    \resizebox{9cm}{!}{\includegraphics{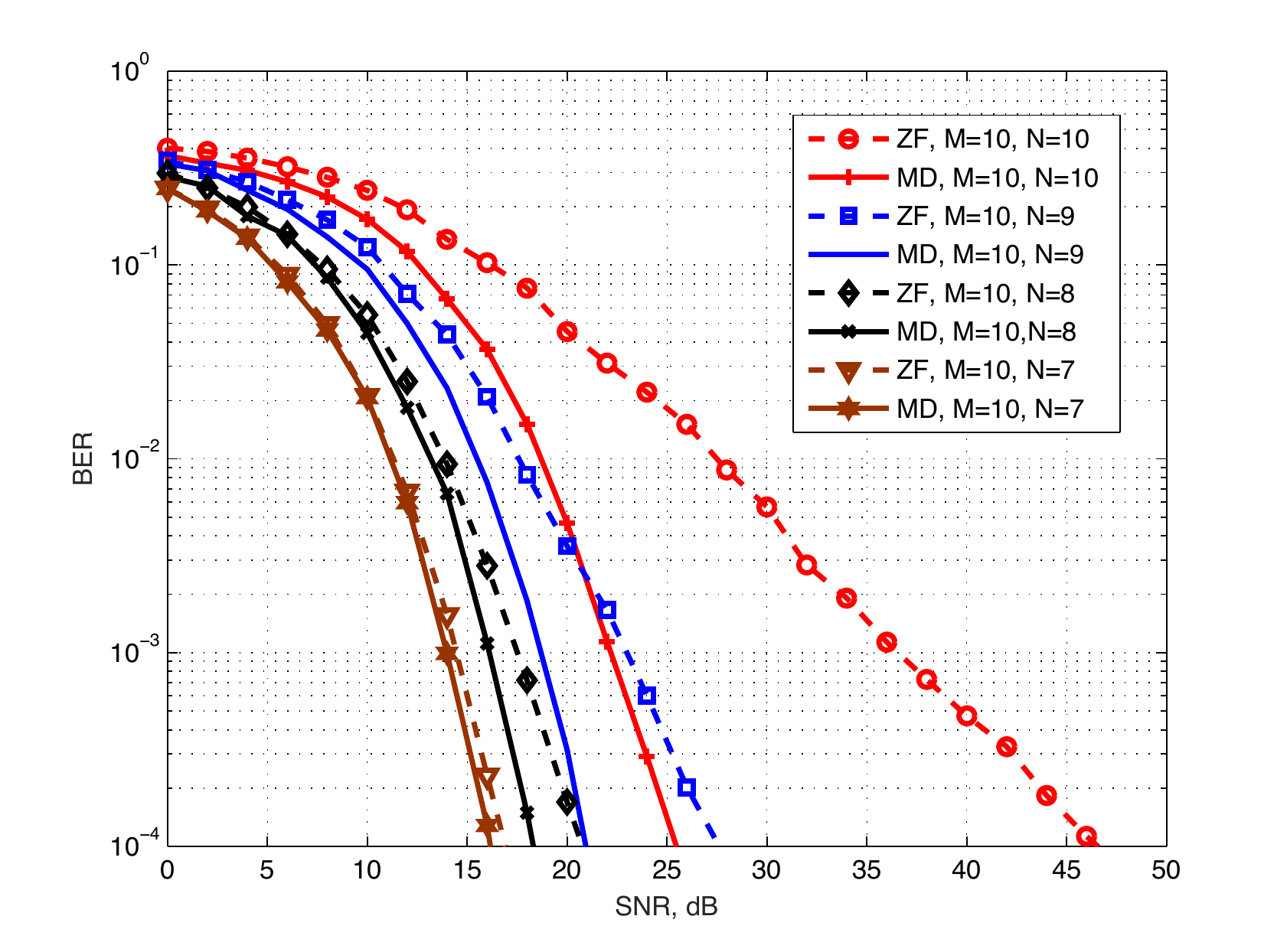}}
    \centering
    \caption{Average BER among all users against SNR, $M=10$, $N=7,8,9,10$ with $\rho=0$.} 
    \label{fig:m10n78910}
\end{figure}
\begin{figure}[ht]
    \centering
    \flushleft
    \resizebox{9cm}{!}{\includegraphics{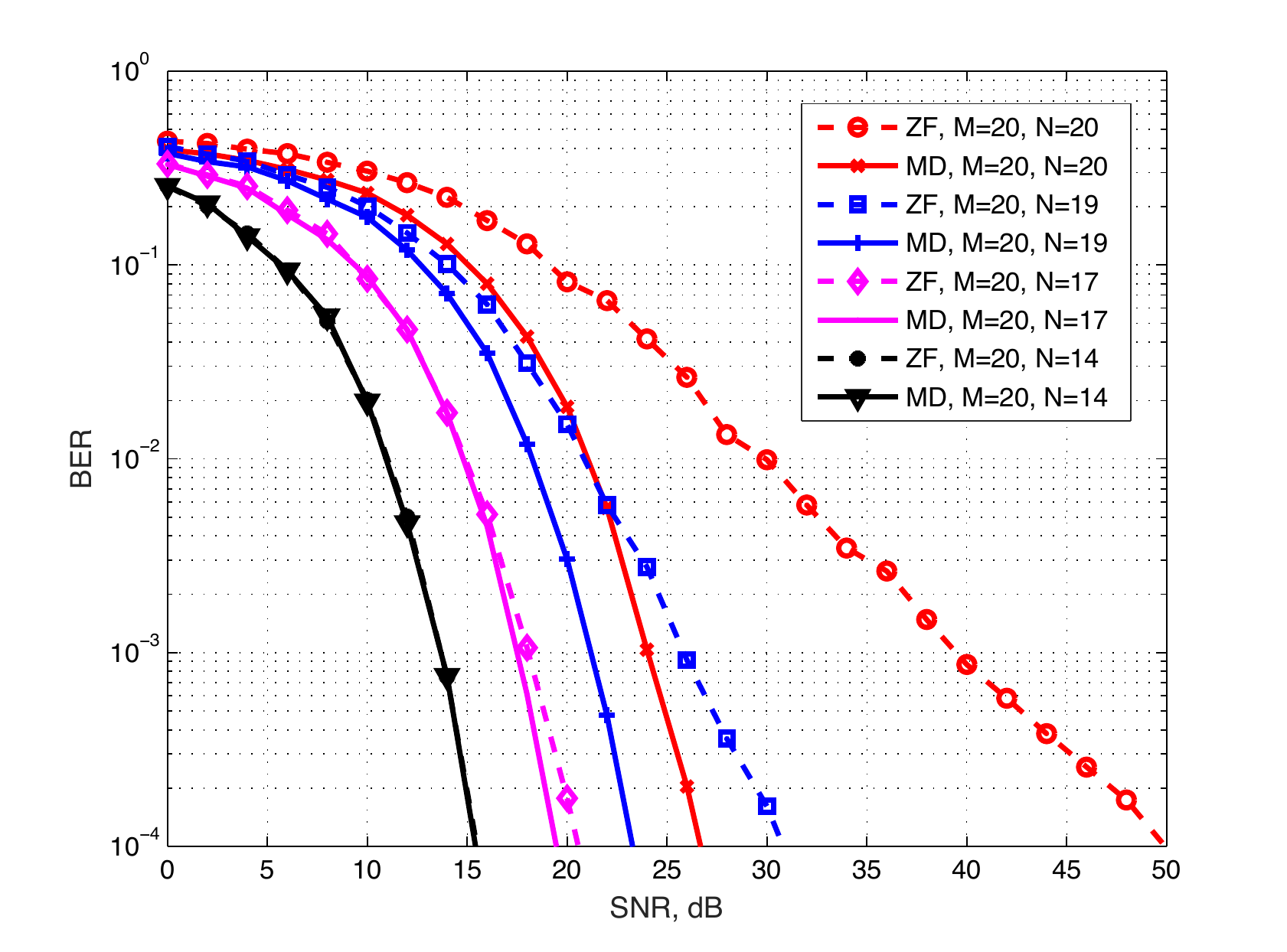}}
    \centering
    \caption{Average BER among all users against SNR, $M=20$, $N=14,17,19, 20$ with $\rho=0$.} 
    \label{fig:MDZFM20N1417190}
\end{figure}
\begin{figure}[ht]
    \centering
    \flushleft
    \resizebox{9cm}{!}{\includegraphics{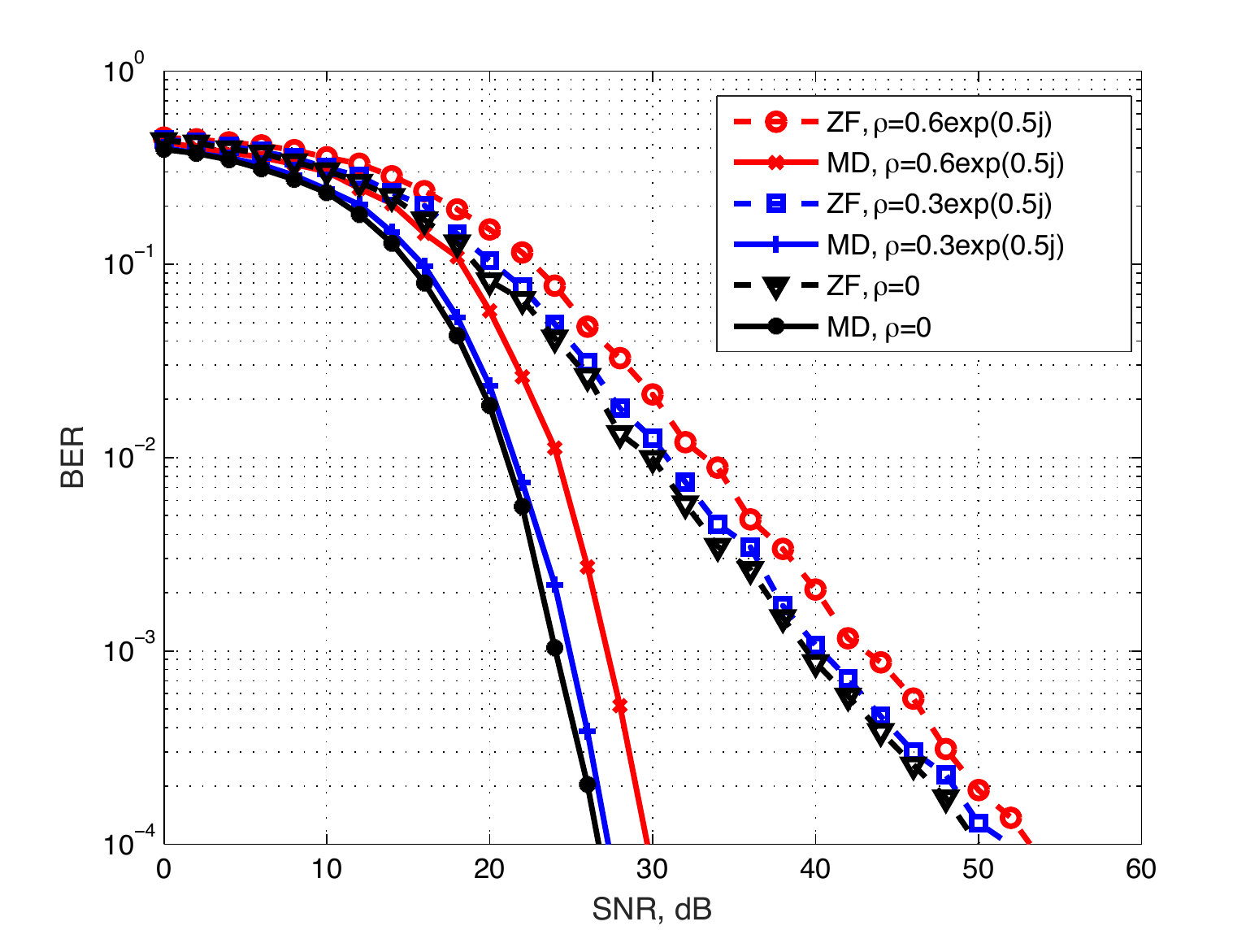}}
    \centering
    \caption{Average BER among all users against SNR, $M=20$, $N=20$ with $\rho=0, 0.3\exp(0.5j)$ and $0.6\exp(0.5j)$.} 
    \label{fig:MDZFM20N20}
\end{figure}
\begin{figure}[ht]
    \centering
    \flushleft
    \resizebox{9cm}{!}{\includegraphics{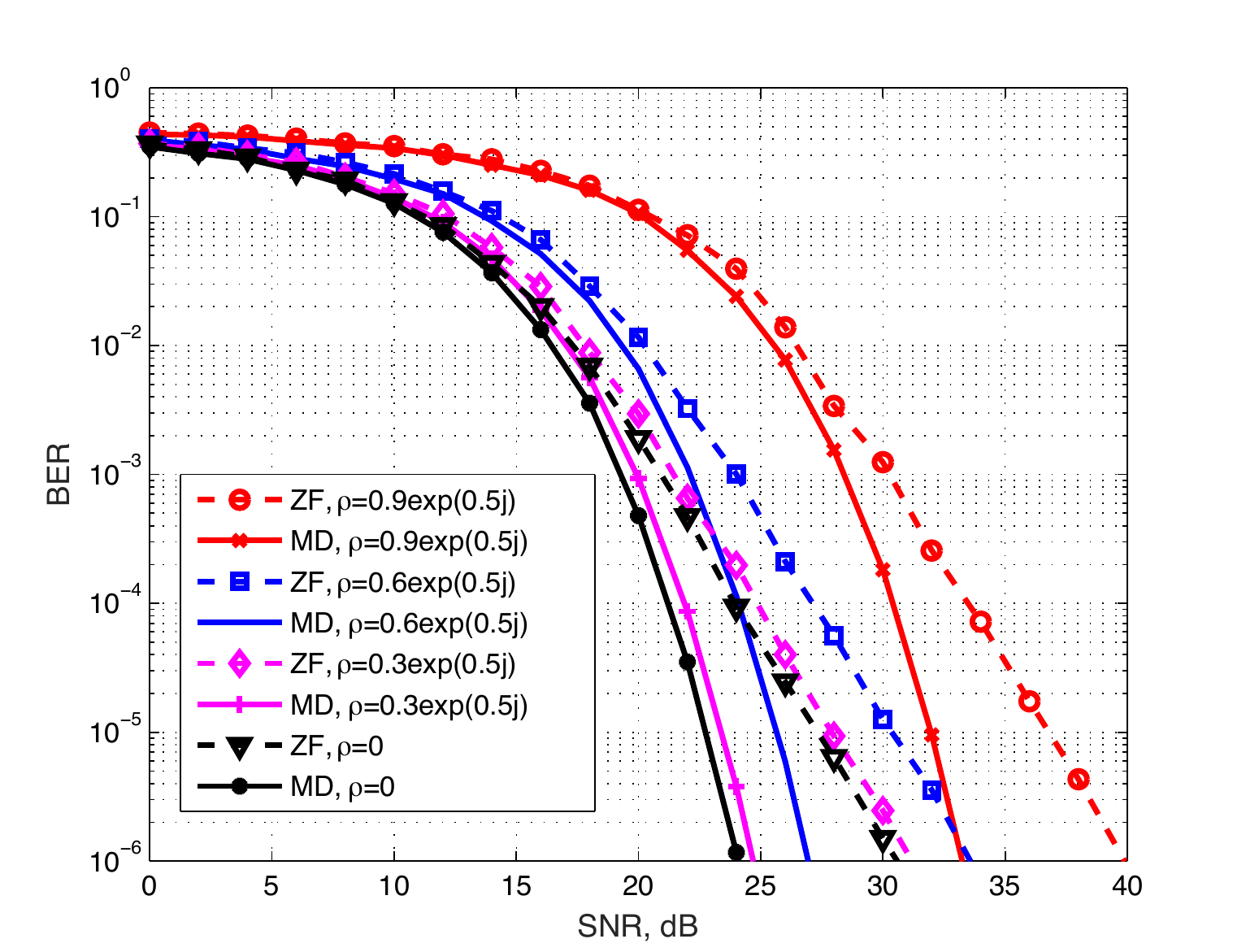}}
    \centering
    \caption{Average BER among all users against SNR, $M=20$, $N=18$ with $\rho=0, 0.3\exp(0.5j)$, $0.6\exp(0.5j)$ and  $0.9\exp(0.5j)$ .} 
    \label{fig:M20N20rho030609}
\end{figure}

\begin{figure}[ht]
    \centering
    \flushleft
    \resizebox{9cm}{!}{\includegraphics{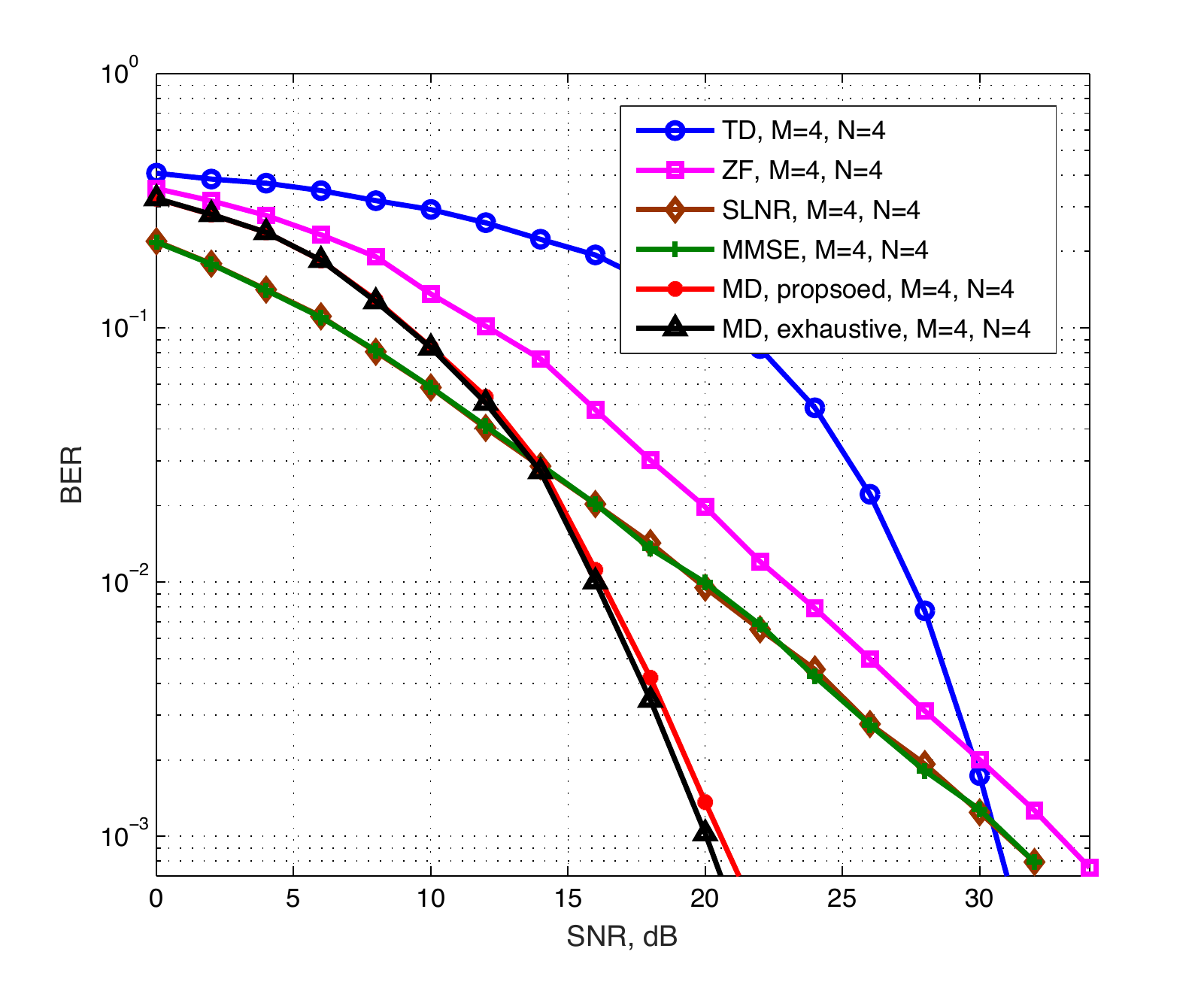}}
    \centering
    \caption{Comparison of the proposed MD method, the exhaustive MD method, ZF, TD, MMSE and SLNR methods with $M=4$ and $N=4$.} 
    \label{fig:ComparisonExhGreedyZFTD}
\end{figure}
\begin{figure}[ht]
    \centering
    \flushleft
    \resizebox{9cm}{!}{\includegraphics{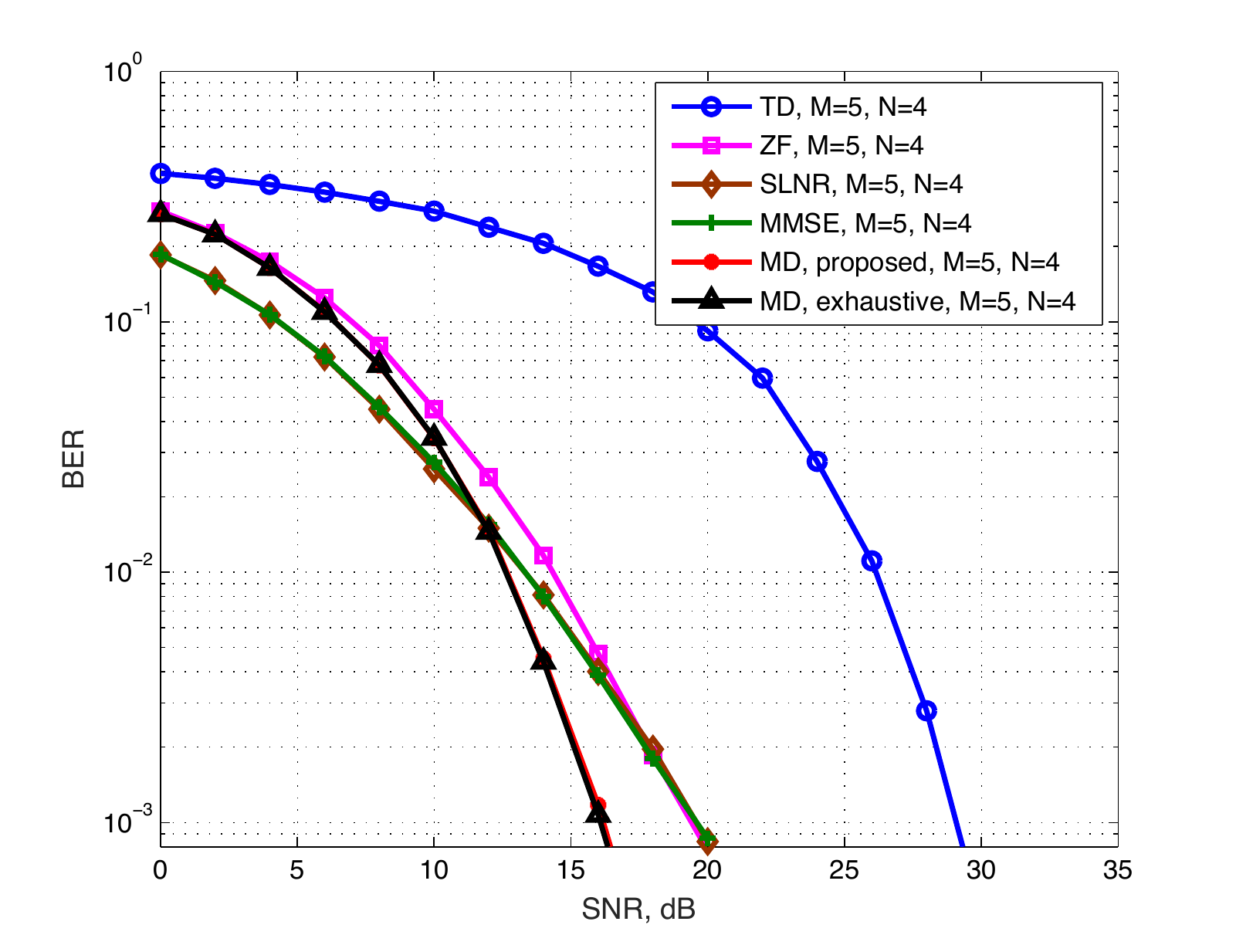}}
    \centering
    \caption{Comparison of the proposed MD method, the exhaustive MD method, ZF, TD, MMSE and SLNR methods with $M=5$ and $N=4$.} 
    \label{fig:ComparisonExhGreedyZFTDM5}
\end{figure}

In addition, to help with our simulation, we assume that the channel covariance matrix ${\bf \Sigma }$ is the commonly-used non-symmetric Kac-Murdock-Szeg$\ddot{\text{o}}$~(KMS) matrix~{\cite{ Loyka01},  
the $(m,n)$-th entry of which is denoted by $\sigma(m-n)$}, i.e., 
\begin{align}\label{kmsmatrix}
\sigma(m-n) =[{\bf \Sigma }]_{mn}  =\left\{ \begin{matrix} 
\rho^{n-m}& m \le n\\
[{\mathbf \Sigma}]^{*}_{nm}&m>n,
\end{matrix} \right.
\end{align}
where $ 0 \le |\rho| < 1$  indicates the degree of correlation. In particular, if $\rho=0$, then ${\mathbf \Sigma}=\mathbf{I}$, i.e., all the entries of $\mathbf{H}$ are i.i.d. Gaussian. Under all these assumptions, we perform five kinds of simulations to test our proposed MD transmission scheme  in terms of uncoded BER. 

The first kind of simulations is to test our MD method for the two user case. Its error performance comparison with the ZF beamforming method and the time division (TD) method is plotted in Fig.~\ref{fig:combined_n2_rayleigh}, where our grouping method for two-user case is examined, with $\rho=0$ and $M=2,4,6$ transmitting antennas, $N=2$ receivers. For the ZF method, each user uses 4-QAM and for the MD and TD methods, the sum-constellation is 16-QAM. It can be observed that the BER performance of the proposed MD method is always better than those of the ZF beamforming and the TD methods. Specifically, the SNR gain at  BER $10^{-4}$ is approximately $15$dB for $M=2$ and $N=2$. However, as the number of transmitter antennas is increased to $M=4$, the BER performance of MD is still better than that of ZF, but the gap between the two methods decreases. Particularly when $M=6$, the performance of our proposed method is almost the same as that of ZF.

The second kind of simulations is to examine that how the correlations among the transmitter antennas affect the error performance of our MD method in the two user situation. The simulation results are shown in Fig.~\ref{fig:m6n2varrho}. It can be seen that the error performance gap between the MD and the ZF method becomes large with $|\rho|$ increasing. In this case with mild correlation, e.g., $\rho=0.3\exp(0.5j)$, the performance gain of our method is not observable. However, when the transmitter antennas are severely correlated, e.g., $\rho=0.9\exp(0.5j)$, our method attains at least $1.5$dB gain at BER $10^{-4}$ over the ZF scheme. 


The third kind of simulations is to test our proposed suboptimal grouping method for the multi-user ($N \ge 3$) MISO BC, as shown in Fig.~\ref{fig:m10n78910}, where $M=10,  N=7,8,9,10$ and the channels are i.i.d. Rayleigh fading. It can be observed that the MD method with user grouping always outperforms the ZF method. Specifically for the case of $M=N=10$, the SNR gain is approximately $20$dB at BER $10^{-4}$. We find that  the closer the number of users $N$ is to the number of transmitters $M$, the larger the performance gap between the proposed MD strategy and ZF method becomes, since when $M$ is close to $N$, there is a higher probability that the Hermitian angle between the channel vector of two users is small and as a consequence, the grouping gain becomes large. A similar conclusion can be drawn from the case with $M=20$, $N=14,17,19,20$, as shown in Fig.~\ref{fig:MDZFM20N1417190}.

Similar to the second kind of simulations, the fourth kind of simulations is to investigate how the channel correlations among the transmitter antennas affect the error performance of our proposed MD-based grouping scheme. The simulation results are shown in Figs.~\ref{fig:MDZFM20N20} and~\ref{fig:M20N20rho030609}.
In Fig.~\ref{fig:MDZFM20N20}, we consider the case with $M=N=20$ and different correlation coefficients.  It can be expected that the BER performance of our proposed method is much better than that of ZF method. In addition, it is not surprising that the BER performance becomes worse when the channel links from the BS becomes more correlated.  Similar observations are also verified for the case with $M=20$ and $N=18$, as shown in Fig.~\ref{fig:M20N20rho030609}.

The last kind of simulation is to compare our proposed MD method with other existing precoding methods in  Figs.~\ref{fig:ComparisonExhGreedyZFTD} and~\ref{fig:ComparisonExhGreedyZFTDM5}. 
Toward this end, in Fig.~\ref{fig:ComparisonExhGreedyZFTD}, we compare the average BER of the MD approach with SLNR based scheme in~\cite{Sayed07} and MMSE method with equal power allocation as well as TD and  ZF methods when $M=4, N=4$. For the MD method, we consider both the proposed suboptimal grouping method described in Algorithm\,\ref{algorithm:suboptgroup} and an exhaustive search grouping scheme that enumerates all the possible grouping methods for seeking the best possible max-min weighted SNR in Problem~\ref{pbm:groupbf}. It can be observed that the TD method has the worst BER in low and moderate SNR regimes. 
It can also be noticed that the SLNR and MMSE methods with equal power allocation have the same BER performance as proved by~\cite{Armour12} and both of them outperform the ZF method.  In addition, we can see that in a low SNR regime, the MMSE and SLNR methods have a lower BER than the MD method. However, in a moderate and high SNR regime, the MD method outperforms all the other methods in terms of BER. Despite the fact that it has  better error performance than the proposed MD approach with the suboptimal grouping method, the exhaustive grouping MD method obtains the marginal  BER gain and costs much higher computational complexity. Therefore, the proposed suboptimal grouping method is greatly desirable in practice. To further demonstrate the error performance comparison of our proposed MD method with other precoding schemes, the scenario with $M=5, N=4$ is given in Fig.~\ref{fig:ComparisonExhGreedyZFTDM5}, where similar conclusion can be drawn.

Here, it should be pointed out clearly that our optimal beamformer design, the optimal grouping scheme and all the resulting simulations are based on the prior assumption that the BS has the knowledge of the perfect CSI. However, in practice, it is difficult to obtain the perfect knowledge of the CSI at BS. Therefore, it would be very necessary to analyze the error performance of our proposed algorithm with the imperfect CSI, especially for the case of multiuser massive MIMO BC. Unfortunately, this problem is too big and too important to have space for any investigation in this paper, but will be further studied in our future research. 

In addition, it also should be mentioned explicitly that in terms of computational complexity, channel state information required and other overheads, 
there are some drawbacks in our proposed MD scheme when compared with  the ZF scheme, as listed in Table~\ref{tab:dis}. 
\begin{table}[hthp]
    \caption{Comparison of MD and ZF Method}
    \begin{center}
    \begin{tabular}{|c||c|c|}
    \hline
     {\rm Aspects}& {\rm MD}& {\rm ZF} \\ \hline\hline
     {\rm Complexity}  &  $\mathcal{O}(M^5)$ &$\mathcal{O}(M^3)$\\ \hline
     {\rm CSI} & CSIT and CSIR & CSIT \\ \hline
     {\rm Overhead} & Grouping Index needed & No additional overhead \\ \hline
    \end{tabular}\label{tab:dis}
    \end{center}
    \end{table}

\section{Conclusion}
In this paper, we have revealed an important  property on PAM and QAM constellations for multiuser communications that any PAM constellation or QAM constellation of large size can be uniquely decomposed into the sum of a group of the scaled version of the PAM or QAM constellations of variety of small flexible sizes. In addition, we have developed two detection algorithms, showing that one of significant advantages of such unique decomposition is that once the sum signal is detected, each individual user signal can be efficiently decoded. Then, we consider a MISO BC with two users. For the special case with two receivers, the optimal beamforming vector is given in a closed-form based on a max-min criterion on the received SNR for the sum-signal. In addition, the SNR gain compared with ZF method is also given explicitly, which can be used to evaluate the effectiveness of our MD method in different channel coefficients.  In the case with more than two receivers, a novel low-complexity grouping transmission scheme based on MD, which aims at improving the condition number of the channel matrix, is proposed, with each group having one or two users. 

Finally, the simulation results have demonstrated that for the Rayleigh channel, if the number of the receivers is far less than the number of BS antennas, our method has the same error performance as the ZF method. However, when the number of users is very close to that of the BS antennas, the error performance of our proposed MD scheme is substantially better than ZF. Moreover, our computer simulations have also shown that when the transmitter antennas are correlated, our presented method is still better than ZF, even if the number of transmitter antennas is larger than that of the receivers.

In conclusion, our QAM MD transmission scheme can be considered as a feasible and concrete approach to the general NOMA method that introduces interference with proper power level superimposed on the desired signal and is possible to be applied to multiuser networks, which would enable a new promising multiple access method.

\section*{Appendix}
\begin{appendices}
\subsection{Proof of Algorithm\,\ref{fastpam}} \label{appendix:algorithm1}For any given $y=g+\xi$, the ML detection of $g$ is exactly equivalent to that of $p=g+\frac{2^K-1}{2}\in\{k\}_{k=0}^{2^K-1}$ from $p+\xi$, which is given by
\begin{eqnarray*}
&&\hat{p}=\left\{
\begin{array}{llll}
0,~~~~~~~~~~~~~~~~~~~~~~~y+\frac{2^K-1}{2}\le0;\\
\lfloor y+\frac{2^K-1}{2}+\frac{1}{2}\rfloor ,0<y+\frac{2^K-1}{2}\le 2^K-1;\\
2^K-1,~~~~~~~~~~~~~~~y+\frac{2^K-1}{2}>2^K-1.
\end{array}
\right.
\end{eqnarray*}
leading us to~\eqref{eqn:sum_ml}.

For $K_i=0$,~\eqref{eqn:user1} and~\eqref{eqn:usern} indeed hold. Therefore, we only consider the case $K_i\neq0$. For presentation simplicity, we define $p_i$  as follows: 1) for any $x_1\in\mathcal{X}_1$, $p_1=x_1+\frac{2^{K_1}-1}{2}$; and 2) for any $x_i\in\mathcal{X}_i$, $p_i=x_i\times2^{-\sum_{\ell=1}^{i-1}K_\ell}+\frac{2^{K_i}-1}{2}$. From the definitions of $\mathcal{X}_i$, we can attain that $p_i\in\{k\}_{k=0}^{2^{K_i}-1}$. In addition, Theorem~\ref{AUDCGPAM} tells us that given $\hat{g}\in\mathcal{G}$ defined by~\eqref{eqn:sum_ml}, $\hat{g}$ can be uniquely decomposed into $\hat{g}=\sum_{i=1}^N\hat{x}_{i}$, where $\hat{x}_i\in\mathcal{X}_i$ and thus, be equivalently rewritten into
 \begin{eqnarray}\label{eqn:audcg_sum}
 \hat{g}=\sum_{i=1}^N\hat{x}_{i}=\hat{p}_1+\sum_{i=2}^N\hat{p}_{i}\times2^{\sum_{\ell=1}^{i-1}K_\ell}-\frac{2^{K}-1}{2}
 \end{eqnarray}
 where $\hat{p}_i\in\{k\}_{k=0}^{2^{K_i}-1}$. It is noticed that for any $\hat{g}\in\mathcal{G}=\{\pm(k-\frac{1}{2})\}_{k=0}^{2^{K-1}}$, we can have $\{\hat{g}+\frac{2^{K}-1}{2}:\hat{g}\in\mathcal{G}\}=\{k\}_{k=0}^{2^{K}-1}$. Then, letting $\hat{p}=\hat{g}+\frac{2^K-1}{2}$ produces an equivalent form of~\eqref{eqn:audcg_sum} as $\hat{p}=\hat{p}_1+\sum_{i=2}^N\hat{p}_{i}\times2^{\sum_{\ell=1}^{i-1}K_\ell}$, where $\hat{p}_i\in\{k\}_{k=0}^{2^{K_i}-1}$. Now, we notice that $\hat{p}_1=\hat{p}\mod 2^{K_1}$ because of the fact that $\sum_{i=2}^N\hat{p}_{i}\times2^{\sum_{\ell=1}^{i-1}K_\ell}\mod 2^{K_1}=0$. This result gives us that $\hat{x}_1=\left(\hat{g}+\frac{2^K-1}{2}\right)\mod 2^{K_1}-\frac{2^{K_1}-1}{2}$, proving~\eqref{eqn:user1}. Also, we observe that $\frac{\hat{p}-\hat{p}_1}{2^{K_1}}=\hat{p}_2+\sum_{i=3}^N\hat{p}_{i}\times2^{\sum_{\ell=2}^{i-1}K_\ell}$ and thus, attain that
 $\hat{p}_2=\frac{\hat{p}-\hat{p}_1}{2^{K_1}}\mod 2^{K_2}$, leading to $\hat{x}_2=\left(\hat{p}_2-\frac{2^{K_1}-1}{2}\right)\times 2^{K_1}$.  Following this process, we can attain that for $2\le i\le N$, $\hat{p}_i=\left(\frac{\hat{p}-\hat{p}\mod2^{\sum_{\ell=1}^{i-1}K_\ell}}{2^{\sum_{\ell=1}^{i-1}K_\ell}}\right)\mod2^{K_i}$. In addition, our notation $\hat{p}_i=\hat{x}_i\times2^{-\sum_{\ell=1}^{i-1}K_\ell}+\frac{2^{K_i}-1}{2}$ allows us to arrive at the desired~\eqref{eqn:usern}.
 
 Therefore, this verifies Algorithm~\ref{fastpam}.\hfill\QED

\subsection{Proof of Theorem~\ref{thm:opttwousr}}\label{appendix:theorem3}
 \emph{Scenario 1: $\mathbf{h}_1 =\tau \mathbf{h}_2, \tau \in \mathbb C$}. If $|\tau|\ge 1$, then, we have $\mathbf{w}^H \mathbf{h}_1 \mathbf{h}_1^H \mathbf{w}- \mathbf{w}^H \mathbf{h}_2 \mathbf{h}_2^H \mathbf{w} =(|\tau|^2 -1|) \mathbf{w}^H \mathbf{h}_2 \mathbf{h}_2^H \mathbf{w} \ge 0$ for any $\mathbf{w}^H\mathbf{w} =P$ and as a result, the original optimization problem degrades into  $\max_{\|\mathbf{w}\|^2 =P}  ~ \mathbf{w}^H \mathbf{h}_2 \mathbf{h}_2^H \mathbf{w}$. Using Cauchy-Schwarz inequality, $\mathbf{w}^{\rm opt} = \frac{\sqrt{P}\mathbf{h}_2}{\|\mathbf{h}_2\|}$ (or $\mathbf{w}^{\rm opt} = \frac{\sqrt{P}\mathbf{h}_1}{\|\mathbf{h}_1\|}$).  The case when $|\tau| <1$ is similar and we omit it here.

\emph{Scenario 2: $\mathbf{h}_1 \neq \tau \mathbf{h}_2, \forall \tau \in \mathbb C$}.
In this case, the overall optimization problem can be divided into the following two problems by introducing restrictions on the original feasible region and the global optimum is the maximum of the two.
\begin{itemize}
\item \emph{Case 1: SNR of user 1 is not worse than that of user 2}
\begin{subequations}\label{optcase1}
\begin{align} \label{optcase1}
&\max_{\mathbf{w}} ~{\mathbf{w}}^H\mathbf{{h}}_2 \mathbf{{h}}_2^H {\mathbf{w}} \\
&{\rm s.t.~~}\mathbf{w}^H \mathbf{A}\mathbf{w}  \ge 0 {\rm ~and~} \mathbf{w}^H\mathbf{w}=P
\end{align}
\end{subequations}
\item \emph{Case 2: SNR of user 2 is better than that of user 1}
\begin{subequations}\label{optcase2}
 \begin{align}
&\max_{\mathbf{w}} ~{\mathbf{w}}^H\mathbf{{h}}_1 \mathbf{{h}}_1^H {\mathbf{w}} \label{case1obj}\\
&{\rm s.t.~~}\mathbf{w}^H \mathbf{A} \mathbf{w}  < 0 {\rm ~and~} \mathbf{w}^H\mathbf{w}=P
\end{align}
\end{subequations}
\end{itemize}
Let us consider \emph{Case 1} first. Using the notation in~\eqref{convertedchannel}, the optimization problem can be reformulated as
\begin{subequations}
\begin{align}
&\max_{\tilde{\mathbf{w}}}~\tilde{\mathbf{w}}^H\mathbf{\tilde{h}}_2 \mathbf{\tilde{h}}_2^H  \tilde{\mathbf{w}} \label{case1:optpbmobj}\\
&{\rm s.t.~~} \lambda_1 |\tilde{w}_1|^2 \ge \lambda_2|\tilde{w}_2|^2 {\rm ~and~} \sum_{\ell=1}^{M} |\tilde{w}_\ell|^2=P
\end{align}
\end{subequations}
 We have the following two observations on the above optimization problem:
\begin{enumerate}
\item From the objective function $\tilde{\mathbf{w}}^H\mathbf{\tilde{h}}_2 \mathbf{\tilde{h}}_2^H  \tilde{\mathbf{w}}=|\sum_{\ell =1}^{M} \tilde{w}_\ell^* \tilde{h}_{2,\ell}|^2$ and the constraint, one optimal choice of the angle of $\tilde{w}_\ell$ is $\arg(\tilde{w}_\ell ) =\arg( \tilde{h}_{2,\ell}), \forall \ell$. 
\item Since $\tilde{\mathbf{h}}_2 = [\tilde{h}_{21},\tilde{h}_{22},0,\ldots, 0]^T$, we should let $|\tilde{w}_\ell|=0,\forall \ell \ge 3$. Otherwise we could let $|\tilde{w}_\ell|=0, \forall \ell \ge 3$ and increase $|\tilde{w}_1|$ and $|\tilde{w}_2|$ slightly without violating the constraint, resulting in an increased objective function.
\end{enumerate}
Hence, the optimization problem can be simplified into
\begin{subequations}\label{optcase1simp}
\begin{align}
&\max_{\tilde{w}_1}~ |\tilde{w}_1| |\tilde{h}_{2,1}|+ \sqrt{P -  |\tilde{w}_1|^2} |\tilde{h}_{2,2}|  \\
&{\rm s.t.~~} \sqrt{\frac{P\lambda_2}{\lambda_1+\lambda_2}}\le |\tilde{w}_1|  \le \sqrt{P}
\end{align}
\end{subequations}
With the help of~\eqref{parameterization}, we can denote the objective function as 
\begin{align*}
g_1( |\tilde{w}_1|) 
= |\tilde{w}_1| \sqrt{\lambda_1}\tan\theta + \sqrt{P -  |\tilde{w}_1|^2}\sqrt{\lambda_2} \sec\theta
\end{align*}
whose derivative is $g_1'( |\tilde{w}_1|) \!\!=\!\sqrt{\lambda_1}\tan\theta - \frac{  |\tilde{w}_1|}{\sqrt{P \!-\!|\tilde{w}_1|^2}}|\sqrt{\lambda_2}\sec\theta$ 
\begin{align*}
g_1'( |\tilde{w}_1|)= \begin{cases}
=0 &|\tilde{w}_1|=\sqrt{\frac{P\lambda_1 }{\lambda_1 +\lambda_2 \csc^2\theta}}\\
>0 &0\le |\tilde{w}_1|<\sqrt{\frac{P\lambda_1 }{\lambda_1 +\lambda_2 \csc^2\theta}}\\
<0 &\sqrt{\frac{P\lambda_1 }{\lambda_1 +\lambda_2 \csc^2\theta}}< |\tilde{w}_1|\le \sqrt{P}
\end{cases}
\end{align*}
Therefore, the solution to problem~\eqref{optcase1simp} is given below:
\begin{enumerate}
\item $0\le \sin \theta \le \frac{\lambda_2}{\lambda_1}$ (i.e., $0\le \sqrt{\frac{P\lambda_1}{\lambda_1 +\lambda_2\csc^2\theta}}\le \sqrt{\frac{P\lambda_2}{\lambda_1+\lambda_2}}$), then
$\max_{\sqrt{\frac{P\lambda_2}{\lambda_1+\lambda_2}}\le |\tilde{w}_1|  \le \sqrt{P}}~g_1( |\tilde{w}_1|) =g_1\Big(\sqrt{\frac{P\lambda_2}{\lambda_1+\lambda_2}}\Big) 
=\sqrt{\frac{P\lambda_1\lambda_2}{\lambda_1+\lambda_2}} \frac{1+\sin\theta}{\cos \theta}$.
\item $\frac{\lambda_2}{\lambda_1}< \sin \theta \le 1$ (i.e., $ \sqrt{\frac{P\lambda_2}{\lambda_1+\lambda_2}} <\sqrt{\frac{P\lambda_1 }{\lambda_1 +\lambda_2\csc^2\theta}} \le  \sqrt{\frac{P\lambda_1}{\lambda_1+\lambda_2}} $), then 
$\max_{\sqrt{\frac{P\lambda_2}{\lambda_1+\lambda_2}}\le |\tilde{w}_1|  \le \sqrt{P}}~g_1\Big( |\tilde{w}_1|\Big) =g_1\Big(\sqrt{\frac{P\lambda_1}{\lambda_1 +\lambda_2 \csc^2\theta}} \Big )
=\frac{\sqrt{P(\lambda_1\sin^2\theta+\lambda_2)}}{\cos \theta}$.
\end{enumerate}
For \emph{Case 2}, by a similar argument on the objective function $\tilde{\mathbf{w}}^H\mathbf{\tilde{h}}_1 \mathbf{\tilde{h}}_1^H  \tilde{\mathbf{w}} =|\sum_{\ell =1}^{M} \tilde{w}_\ell^* \tilde{h}_{1,\ell}|^2$, we have $\arg(\tilde{w}_\ell ) =\arg( \tilde{h}_{1,\ell}), \forall \ell$, and $|\tilde{w}_\ell|=0, \forall \ell \ge 3$. The resulting optimization problem can be reformulated as
\begin{subequations}\label{optcase2simp}
 \begin{align}
&\max_{\tilde{w}_1}~ |\tilde{w}_1| |\tilde{h}_{1,1}|+ \sqrt{P -  |\tilde{w}_1|^2} |\tilde{h}_{1,2}|  \\
&{\rm s.t.~~}  0 \le  |\tilde{w}_1| \le \sqrt{\frac{P\lambda_2}{\lambda_1+\lambda_2}}
\end{align}
\end{subequations}
With the aid of~\eqref{parameterization}, we can represent the objective function as $g_2( |\tilde{w}_1|) = |\tilde{w}_1| \sqrt{\lambda_1}\sec\theta+ \sqrt{P-  |\tilde{w}_1|^2} \sqrt{\lambda_2} \tan \theta$. Since its first order derivative is 
$g_2'( |\tilde{w}_1|)=  \sqrt{\lambda_1}\sec\theta - \frac{  |\tilde{w}_1|}{\sqrt{P - |\tilde{w}_1|^2}} \sqrt{\lambda_2} \tan \theta$, we have
\begin{align*}
g_2'( |\tilde{w}_1|)= \begin{cases}
=0 &|\tilde{w}_1|=\sqrt{\frac{P\lambda_1}{\lambda_1+\lambda_2\sin^2 \theta }}\\
>0 &0\le |\tilde{w}_1|<\sqrt{\frac{P\lambda_1}{\lambda_1+\lambda_2\sin^2 \theta }}\\
<0 &\sqrt{\frac{P\lambda_1}{\lambda_1+\lambda_2\sin^2 \theta }}< |\tilde{w}_1|\le1
\end{cases}
\end{align*}
Therefore, the solution to~\eqref{optcase2simp} can be determined as follows:
\begin{enumerate}
\item $0\le \sin \theta < \frac{\lambda_1}{\lambda_2}$ (i.e., $\sqrt{\frac{P\lambda_2}{\lambda_1+\lambda_2}} <\sqrt{\frac{P\lambda_1}{\lambda_1+\lambda_2\sin^2 \theta }}\le \sqrt{P}$), then
$\max_{0 \le  |\tilde{w}_1| \le \sqrt{\frac{P\lambda_2}{\lambda_1+\lambda_2}}}~g_2\Big( |\tilde{w}_1|\Big) =g_2\Big( \sqrt{\frac{P\lambda_2}{\lambda_1+\lambda_2}}\Big)
=\sqrt{\frac{P\lambda_1\lambda_2}{\lambda_1+\lambda_2}} \frac{1+\sin\theta}{\cos \theta}$.

\item $\frac{\lambda_1}{\lambda_2}\le \sin \theta \le 1$ (i.e., $0< \sqrt{\frac{P\lambda_1}{\lambda_1+\lambda_2\sin^2 \theta }} \le \sqrt{\frac{P\lambda_2}{\lambda_1+\lambda_2}}$), then 
$\max_{0 \le  |\tilde{w}_1| \le \sqrt{\frac{P\lambda_2}{\lambda_1+\lambda_2}}}~g_2( |\tilde{w}_1|) =g_2\Big(\sqrt{\frac{P\lambda_1}{\lambda_1+\lambda_2\sin^2 \theta }}\Big)= \frac{\sqrt{P(\lambda_1+\lambda_2\sin^2 \theta)}}{\cos \theta} $.
\end{enumerate}
The overall maximum value of the original problem is the maximum of the two cases.

\subsection{Proof of Lemma~\ref{lemma:eigchn}}\label{appendix:lemma1}
Let the singular value decomposition (SVD) of $\mathbf{H}$ be $\mathbf{H}=\mathbf{U}_1 \mathbf{\Sigma}_1 \mathbf{V}_1^H$, where $\mathbf{U}_1 \in \mathbb{C}^{M\times 2}$ is a tall column-wise unitary matrix, $\mathbf{V}_1=[\mathbf{v}_1 ~ \mathbf{v}_2]=\Bigg[\begin{matrix} v_{1,1}&v_{1,2}\\ v_{2,1} &v_{2,2} \end{matrix}\Bigg] \in \mathbb{C}^{2\times 2}$ is a unitary matrix, and $\mathbf{\Sigma}_1={\rm diag}(\sqrt{\mu}_1, \sqrt{\mu}_2)$. 
Then, we have $[\mathbf{h}_1 ~-\mathbf{h}_2]=\mathbf{U}_1 \mathbf{\Sigma}_1 \Bigg[\begin{matrix} v_{1,1}^*&-v_{2,1}^*\\ v_{1,2}^* &-v_{2,2}^* \end{matrix}\Bigg]$. 
and thus, equation~\eqref{eqn:amatrix} can be represented by $\mathbf{A} =[\mathbf{h}_1 ~-\mathbf{h}_2][\mathbf{h}_1 ~\mathbf{h}_2]^H = \mathbf{U}_1 \mathbf{B} \mathbf{U}_1^H$, where 
\begin{align}\label{eqn:matrixb}
\mathbf{B}= \mathbf{\Sigma}_1 \Bigg[\begin{matrix} |v_{1,1}|^2 - |v_{2,1}|^2& v_{1,1}^* v_{1,2} - v_{2,1}^* v_{2,2}\\ v_{1,2}^* v_{1,1}-v_{2,2}^* v_{2,1} &  |v_{1,2}|^2 - |v_{2,2}|^2 \end{matrix}\Bigg]\mathbf{\Sigma}_1.
\end{align}
Notice that matrix $\mathbf{A}$ and $\mathbf{B}$ have the same non-zero eigenvalues, i.e.,
${\rm diag}(\lambda_1, -\lambda_2)={\rm eig}( \mathbf{B})$.
Hence, in order to use $a, b$ and $c$ to represent $\lambda_1, \lambda_2, \mu_1$ and $\mu_2$, we need to calculate $\mathbf{V}_1$. Since $\|\mathbf{h}_1\|^2 =a, \|\mathbf{h}_2\|^2 =b$ and $|\mathbf{h}_1^H \mathbf{h}_2|=c$, if we let $\arg (\mathbf{h}_1^H \mathbf{h}_2 )=\phi_c$, then, we have 
\begin{align}\label{eqn:chneigmtx}
\mathbf{H}^H\mathbf{H} &=\mathbf{V}_1 \mathbf{\Sigma}_1^2 \mathbf{V}_1^H=\begin{bmatrix}a & ce^{j \phi_c}\\
ce^{-j \phi_c} & b\end{bmatrix}.
\end{align}
Hence, $\mathbf{V}_1$ is the eigenvector matrix of $\mathbf{H}^H\mathbf{H}$ and $\mu_1$ and $\mu_2$ are its eigenvalues, which must
satisfy the following characteristic equation,
$\Bigg|\begin{matrix}a -\mu_i& ce^{j \phi_c}\\
ce^{-j \phi_c} & b-\mu_i\end{matrix}\Bigg|=0$, for $i=1, 2$. Therefore, we have $\mu_1 =\frac{a+b+\sqrt{ (a-b)^2 +4c^2}}{2}$, $\mu_2 = \frac{a+b-\sqrt{ (a-b)^2 +4c^2}}{2}$.

Correspondingly, the two column vectors of $\mathbf{V}_1=[\mathbf{v}_1 ~ \mathbf{v}_2]$ must satisfy the following equations:
\begin{align*}
\begin{bmatrix}a -\mu_1& ce^{j \phi_c}\\
ce^{-j \phi_c} & b -\mu_1\end{bmatrix} \mathbf{v}_1 =\mathbf{0}, \quad 
\begin{bmatrix}a -\mu_2& ce^{j \phi_c}\\
ce^{-j \phi_c} & b-\mu_2\end{bmatrix}\mathbf{v}_2 =\mathbf{0}.
\end{align*} 
Let $d=\sqrt{(a-b)^2+4c^2}$. Then, we have 
\begin{align*}
\mathbf{v}_1 &= \bigg[\frac{-2c e^{j \phi_c}}{\sqrt{2d^2 -2(a-b)d}}~ \frac{a-b-d}{\sqrt{2d^2 -2(a-b)d  }}\bigg]^T, \\
\mathbf{v}_2 &= \bigg[\frac{-2ce^{j \phi_c}}{\sqrt{2d^2 +2(a-b)d}}~ \frac{a-b+d}{\sqrt{2d^2 +2(a-b)d  }}\bigg]^T,
\end{align*}
and $\mu_1 =\frac{a+b+d}{2}$, $\mu_2= \frac{a+b-d}{2}$ and 
$\sqrt{\mu_1\mu_2} =\sqrt{ab-c^2}$. Hence,~\eqref{eqn:matrixb} can be expressed by

\begin{align*}
\mathbf{B} 
&=\frac{1}{2d}\begin{bmatrix}(a-b)(a+b+d) &4c\sqrt{ab-c^2}\\
4c\sqrt{ab-c^2} & -(a-b)(a+b-d)  \end{bmatrix}.
\end{align*}
On the other hand, the eigenvalues of $\mathbf{A}$, i.e., $\lambda_1, -\lambda_2$ must satisfy the following characteristic equation:
\begin{align}
\Bigg|\begin{matrix} \frac{(a-b)(a+b+d)}{2d}  -\lambda& \frac{4c\sqrt{ab-c^2}}{2d}\\
\frac{4c\sqrt{ab-c^2}}{2d} & \frac{-(a-b)(a+b-d)}{2d}  -\lambda\end{matrix}\Bigg|=0.
\end{align}
Solving this equation and after some algebraic manipulations, we attain
$\lambda_1 = \frac{a-b+\sqrt{(a+b)^2 - 4c^2}  }{2}$,
$\lambda_2 = \frac{-a+b +\sqrt{(a+b)^2 - 4c^2} }{2}$.

This completes the proof of Lemma~\ref{lemma:eigchn}.
\hfill$\Box$

\end{appendices}
\small
\bibliographystyle{ieeetr}
\bibliography{tzzt}

\begin{thebibliography}{10}

\bibitem{Cover72}
T.~Cover, ``Broadcast channels,'' {\em IEEE Trans. Inf. Theory}, vol.~18,
  pp.~2--14, Jan. 1972.

\bibitem{Bergmans73}
P.~Bergmans, ``Random coding theorem for broadcast channels with degraded
  components,'' {\em IEEE Trans. Inf. Theory}, vol.~19, pp.~197--207, Mar.
  1973.

\bibitem{Gallager74}
R.~G. Gallager, ``Coding and capacity for degraded broadcast channels,'' {\em
  Probl. Peredachi Inf.}, vol.~10, pp.~185--193, Oct. 1974.

\bibitem{Meulen1977}
E.~van~der Meulen, ``A survey of multi-way channels in information theory:
  1961-1976,'' {\em IEEE Trans. Inf. Theory}, vol.~23, pp.~1--37, Jan. 1977.

\bibitem{Cover98}
T.~Cover, ``Comments on broadcast channels,'' {\em IRE Trans. Inf. Theory},
  vol.~44, pp.~2524--2530, Oct. 1998.

\bibitem{Kim12}
A.~E. Gamal and Y.-H. Kim, {\em Network Information Theory}.
\newblock Cambridge Univ. Press, 2012.

\bibitem{telatar95}
I.~Telatar, ``Capacity of multiple antenna {G}aussian channels,'' {\em Europ.
  Trans. Telecommu.}, vol.~10, pp.~585--595, Nov.-Dec. 1999.

\bibitem{foschini98}
G.~Foschini and M.~Gans, ``On limits of wireless communications in a fading
  environment when using multiple antenna,'' {\em Wireless Personal Commun.},
  vol.~6, pp.~311--335, Mar. 1998.

\bibitem{Caire03}
G.~Caire and S.~Shamai, ``On the achievable throughput of a multiantenna
  {G}aussian broadcast channel,'' {\em IEEE Trans. Inf. Theory}, vol.~49,
  pp.~1691--1706, Jul. 2003.

\bibitem{Costa83}
M.~Costa, ``Writing on dirty paper,'' {\em IEEE Trans. Inf. Theory}, vol.~29,
  pp.~439--441, May 1983.

\bibitem{Goldsmith03}
S.~Vishwanath, N.~Jindal, and A.~Goldsmith, ``Duality, achievable rates, and
  sum-rate capacity of {G}aussian {MIMO} broadcast channels,'' {\em IEEE Trans.
  Inf. Theory}, vol.~49, pp.~2658--2668, Oct. 2003.

\bibitem{Tse03}
P.~Viswanath and D.~Tse, ``Sum capacity of the vector {G}aussian broadcast
  channel and uplink-downlink duality,'' {\em IEEE Trans. Inf. Theory},
  vol.~49, pp.~1912--1921, Aug. 2003.

\bibitem{Wei04}
W.~Yu and J.~Cioffi, ``Sum capacity of {G}aussian vector broadcast channels,''
  {\em IEEE Trans. Inf. Theory}, vol.~50, pp.~1875--1892, Sep. 2004.

\bibitem{Weingarten06}
H.~Weingarten, Y.~Steinberg, and S.~Shamai, ``The capacity region of the
  {G}aussian multiple-input multiple-output broadcast channel,'' {\em IEEE
  Trans. Inf. Theory}, vol.~52, pp.~3936--3964, Sep. 2006.

\bibitem{Rajan09}
N.~Deshpande and B.~S. Rajan, ``Constellation constrained capacity of two-user
  broadcast channels,'' in {\em Proc. IEEE Global Commun. Conf. (GLOBECOM'09)},
  pp.~1--6, Nov. 2009.

\bibitem{Knopp10}
R.~Ghaffar and R.~Knopp, ``Near optimal linear precoder for multiuser {MIMO}
  for discrete alphabets,'' in {\em Proc. IEEE Int. Conf. Commun. (ICC'10)},
  pp.~1--5, May 2010.

\bibitem{Xiao11}
C.~Xiao, Y.~R. Zheng, and Z.~Ding, ``Globally optimal linear precoders for
  finite alphabet signals over complex vector {G}aussian channels,'' {\em IEEE
  Trans. Signal Process.}, vol.~59, pp.~3301--3314, Jul. 2011.

\bibitem{Gao12}
Y.~Wu, M.~Wang, C.~Xiao, Z.~Ding, and X.~Gao, ``Linear precoding for {MIMO}
  broadcast channels with finite-alphabet constraints,'' {\em IEEE Trans.
  Wireless Commun.}, vol.~11, pp.~2906--2920, Aug. 2012.

\bibitem{Tomlinson71}
M.~Tomlinson, ``New automatic equaliser employing modulo arithmetic,'' {\em
  Electron. Lett.}, vol.~7, pp.~138--139, Mar. 1971.

\bibitem{Windpassinger04}
C.~Windpassinger, R.~Fischer, T.~Vencel, and J.~Huber, ``Precoding in
  multiantenna and multiuser communications,'' {\em IEEE Trans. Wireless
  Commun.}, vol.~3, pp.~1305--1316, Jul. 2004.

\bibitem{Garcia14}
A.~Garcia-Rodriguez and C.~Masouros, ``Power-efficient {T}omlinson-{H}arashima
  precoding for the downlink of multi-user {MISO} systems,'' {\em IEEE Trans.
  Commun.}, vol.~62, pp.~1884--1896, Jun. 2014.

\bibitem{Viswanathan03}
H.~Viswanathan, S.~Venkatesan, and H.~Huang, ``Downlink capacity evaluation of
  cellular networks with known-interference cancellation,'' {\em IEEE J. Sel.
  Areas Commun.}, vol.~21, pp.~802--811, Jun. 2003.

\bibitem{Cioffi05}
W.~Yu, D.~Varodayan, and J.~Cioffi, ``Trellis and convolutional precoding for
  transmitter-based interference presubtraction,'' {\em IEEE Trans. Commun.},
  vol.~53, pp.~1220--1230, Jul. 2005.

\bibitem{Liu98}
F.~Rashid-Farrokhi, K.~Liu, and L.~Tassiulas, ``Transmit beamforming and power
  control for cellular wireless systems,'' {\em IEEE J. Sel. Areas Commun.},
  vol.~16, pp.~1437--1450, Oct. 1998.

\bibitem{Schubert02}
M.~Schubert and H.~Boche, ``Joint `dirty paper' pre-coding and downlink
  beamforming,'' in {\em Proc. IEEE 7th International Symposium on Spread
  Spectrum Techniques and Applications,}, vol.~2, pp.~536--540 vol.2, 2002.

\bibitem{Schubert04}
M.~Schubert and H.~Boche, ``Solution of the multiuser downlink beamforming
  problem with individual {SINR} constraints,'' {\em IEEE Trans. Veh.
  Technol.}, vol.~53, pp.~18--28, Jan. 2004.

\bibitem{Hochwald05}
C.~Peel, B.~Hochwald, and A.~Swindlehurst, ``A vector-perturbation technique
  for near-capacity multiantenna multiuser communication-part {I}: channel
  inversion and regularization,'' {\em IEEE Trans. Commun.}, vol.~53,
  pp.~195--202, Jan. 2005.

\bibitem{Sayed07}
M.~Sadek, A.~Tarighat, and A.~H. Sayed, ``A leakage-based precoding scheme for
  downlink multi-user {MIMO} channels,'' {\em IEEE Trans. Wireless Commun.},
  vol.~6, pp.~1711--1721, May 2007.

\bibitem{Shamai08}
A.~Wiesel, Y.~Eldar, and S.~Shamai, ``Zero-forcing precoding and generalized
  inverses,'' {\em IEEE Trans. Signal Process.}, vol.~56, pp.~4409--4418, Sep.
  2008.

\bibitem{Spencer04}
Q.~Spencer, A.~Swindlehurst, and M.~Haardt, ``Zero-forcing methods for downlink
  spatial multiplexing in multiuser {MIMO} channels,'' {\em IEEE Trans. Signal
  Process.}, vol.~52, pp.~461--471, Feb. 2004.

\bibitem{Samardzija03}
D.~Samardzija and N.~Mandayam, ``Multiple antenna transmitter optimization
  schemes for multiuser systems,'' in {\em Proc. IEEE 58th Vehi. Tech. Conf.
  (VTC Fall'03)}, vol.~1, pp.~399--403 Vol.1, Oct. 2003.

\bibitem{Kobayashi81}
T.~S. Han and K.~Kobayashi, ``A new achievable rate region for the interference
  channel,'' {\em IEEE Trans. Inf. Theory}, vol.~27, pp.~49--60, Jan. 1981.

\bibitem{Cadambe08}
V.~R. Cadambe and S.~A. Jafar, ``Interference alignment and degrees of freedom
  of the {K}-user interference channel,'' {\em IEEE Trans. Inf. Theory},
  vol.~54, pp.~3425--3441, Aug. 2008.

\bibitem{Bresler10}
G.~Bresler, A.~Parekh, and D.~N.~C. Tse, ``The approximate capacity of the
  many-to-one and one-to-many {G}aussian interference channels,'' {\em IEEE
  Trans. Inf. Theory}, vol.~56, pp.~4566--4592, Sep. 2010.

\bibitem{Yihongwu15}
Y.~Wu, S.~S. Shitz, and S.~Verd{\'u}, ``Information dimension and the degrees
  of freedom of the interference channel,'' {\em IEEE Trans. Inf. Theory},
  vol.~61, pp.~256--279, Jan. 2015.

\bibitem{YiLiu15}
Y.~Liu, Y.~Zhang, R.~Yu, and S.~Xie, ``Integrated energy and spectrum
  harvesting for {5G} wireless communications,'' {\em IEEE Network}, vol.~29,
  pp.~75--81, May 2015.

\bibitem{Ganzheng14mag}
G.~Zheng, I.~Krikidis, C.~Masouros, S.~Timotheou, D.~A. Toumpakaris, and
  Z.~Ding, ``Rethinking the role of interference in wireless networks,'' {\em
  IEEE Commun. Mag.}, vol.~52, pp.~152--158, Nov. 2014.

\bibitem{Masouros13}
C.~Masouros, T.~Ratnarajah, M.~Sellathurai, C.~B. Papadias, and A.~K. Shukla,
  ``Known interference in the cellular downlink: a performance limiting factor
  or a source of green signal power?,'' {\em IEEE Commun. Mag.}, vol.~51,
  pp.~162--171, Oct. 2013.

\bibitem{Masouros09}
C.~Masouros and E.~Alsusa, ``Dynamic linear precoding for the exploitation of
  known interference in {MIMO} broadcast systems,'' {\em IEEE Trans. Wireless
  Commun.}, vol.~8, pp.~1396--1404, Mar. 2009.

\bibitem{Masouros15}
C.~Masouros and G.~Zheng, ``Exploiting known interference as green signal power
  for downlink beamforming optimization,'' {\em IEEE Trans. Signal Process.},
  vol.~63, pp.~3628--3640, Jul. 2015.

\bibitem{Masouros12feb}
C.~Masouros and T.~Ratnarajah, ``Interference as a source of green signal power
  in cognitive relay assisted co-existing {MIMO} wireless transmissions,'' {\em
  IEEE Trans. Commun.}, vol.~60, pp.~525--536, Feb. 2012.

\bibitem{Ruizhang15}
Y.~Zeng and R.~Zhang, ``Optimized training design for wireless energy
  transfer,'' {\em IEEE Trans. Commun.}, vol.~63, pp.~536--550, Feb. 2015.

\bibitem{Zhuhan15}
X.~Lu, P.~Wang, D.~Niyato, D.~I. Kim, and Z.~Han, ``Wireless networks with {RF}
  energy harvesting: A contemporary survey,'' {\em IEEE Commun. Surveys Tuts.},
  vol.~17, no.~2, pp.~757--789, 2015.

\bibitem{Krikidis14}
I.~Krikidis, S.~Timotheou, S.~Nikolaou, G.~Zheng, D.~W.~K. Ng, and R.~Schober,
  ``Simultaneous wireless information and power transfer in modern
  communication systems,'' {\em IEEE Commun. Mag.}, vol.~52, pp.~104--110, Nov.
  2014.

\bibitem{Schober13}
D.~W.~K. Ng, E.~S. Lo, and R.~Schober, ``Wireless information and power
  transfer: Energy efficiency optimization in {OFDMA} systems,'' {\em IEEE
  Trans. Wireless Commun.}, vol.~12, pp.~6352--6370, Dec. 2013.

\bibitem{Ruizhang13}
R.~Zhang and C.~K. Ho, ``{MIMO} broadcasting for simultaneous wireless
  information and power transfer,'' {\em IEEE Trans. Wireless Commun.},
  vol.~12, pp.~1989--2001, May 2013.

\bibitem{Kaibinhuang14}
K.~Huang and V.~K.~N. Lau, ``Enabling wireless power transfer in cellular
  networks: Architecture, modeling and deployment,'' {\em IEEE Trans. Wireless
  Commun.}, vol.~13, pp.~902--912, Feb. 2014.

\bibitem{Zhiguoding14}
Z.~Ding, I.~Krikidis, B.~Sharif, and H.~V. Poor, ``Wireless information and
  power transfer in cooperative networks with spatially random relays,'' {\em
  IEEE Trans. Wireless Commun.}, vol.~13, pp.~4440--4453, Aug. 2014.

\bibitem{Hossain14}
E.~Hossain, M.~Rasti, H.~Tabassum, and A.~Abdelnasser, ``Evolution toward {5G}
  multi-tier cellular wireless networks: An interference management
  perspective,'' {\em IEEE Wireless Commun.}, vol.~21, pp.~118--127, Jun. 2014.

\bibitem{Ganzheng13}
G.~Zheng, I.~Krikidis, J.~Li, A.~P. Petropulu, and B.~Ottersten, ``Improving
  physical layer secrecy using full-duplex jamming receivers,'' {\em IEEE
  Trans. Signal Process.}, vol.~61, pp.~4962--4974, Oct. 2013.

\bibitem{Shu76}
T.~Kasami and S.~Lin, ``Coding for a multiple-access channel,'' {\em IEEE
  Trans. Inf. Theory}, vol.~22, pp.~129--137, Mar. 1976.

\bibitem{Shu78}
T.~Kasami and S.~Lin, ``Bounds on the achievable rates of block coding for a
  memoryless multiple-access channel,'' {\em IEEE Trans. Inf. Theory}, vol.~24,
  pp.~187--197, Mar. 1978.

\bibitem{Tilborg85}
P.~van~den Braak and H.~van Tilborg, ``A family of good uniquely decodable code
  pairs for the two-access binary adder channel,'' {\em IEEE Trans. Inf.
  Theory}, vol.~31, pp.~3--9, Jan. 1985.

\bibitem{Ahlswede99}
R.~Ahlswede and V.~Balakirsky, ``Construction of uniquely decodable codes for
  the two-user binary adder channel,'' {\em IEEE Trans. Inf. Theory}, vol.~45,
  pp.~326--330, Jan. 1999.

\bibitem{Chevillat81}
P.~Chevillat, ``N-user trellis coding for a class of multiple-access
  channels,'' {\em IEEE Trans. Inf. Theory}, vol.~27, pp.~114--120, Jan. 1981.

\bibitem{Yoshida97}
H.~Murata and S.~Yoshida, ``Trellis-coded cochannel interference canceller for
  microcellular radio,'' {\em IEEE Trans. Commun.}, vol.~45, pp.~1088--1094,
  Sep. 1997.

\bibitem{Yoshida98}
Y.~Li, H.~Murata, and S.~Yoshida, ``Coding for multi-user detection in
  interference channel,'' in {\em Proc. IEEE Global Commun. Conf.
  (GLOBECOM'98)}, vol.~6, pp.~3596--3601 vol.6, 1998.

\bibitem{Yongacoglu04}
W.~Zhang, C.~D'Amours, and A.~Yongacoglu, ``Trellis coded modulation design for
  multi-user systems on {AWGN} channels,'' in {\em Proc. IEEE 59th Vehi. Tech.
  Conf. (VTC Spring'04)}, vol.~3, pp.~1722--1726 Vol.3, May 2004.

\bibitem{Vetterli93}
K.~Ramchandran, A.~Ortega, K.~Uz, and M.~Vetterli, ``Multiresolution broadcast
  for digital {HDTV} using joint source/channel coding,'' {\em IEEE J. Sel.
  Areas Commun.}, vol.~11, pp.~6--23, Jan. 1993.

\bibitem{leefangwei93}
L.-F. Wei, ``Coded modulation with unequal error protection,'' {\em IEEE Trans.
  Commun.}, vol.~41, pp.~1439--1449, Oct. 1993.

\bibitem{Bhargava07}
J.~Hossain, M.-S. Alouini, and V.~Bhargava, ``Multi-user opportunistic
  scheduling using power controlled hierarchical constellations,'' {\em IEEE
  Trans. Wireless Commun.}, vol.~6, pp.~1581--1586, May 2007.

\bibitem{malladi2012}
D.~Malladi, ``Hierarchical modulation for communication channels in
  single-carrier frequency division multiple access,'' Sep. 2012.
\newblock US Patent 8,259,848.

\bibitem{Harshan11}
J.~Harshan and B.~Rajan, ``On two-user {G}aussian multiple access channels with
  finite input constellations,'' {\em IEEE Trans. Inf. Theory}, vol.~57,
  pp.~1299--1327, Mar. 2011.

\bibitem{Saito14}
Y.~Saito, Y.~Kishiyama, A.~Benjebbour, T.~Nakamura, A.~Li, and K.~Higuchi,
  ``Non-orthogonal multiple access (noma) for cellular future radio access,''
  in {\em Proc. IEEE 77th Vehi. Tech. Conf. (VTC Spring'13)}, pp.~1--5, June
  2013.

\bibitem{Zhiguoding15}
Z.~Ding, Y.~Liu, J.~Choi, Q.~Sun, M.~Elkashlan, C.~I, and H.~V. Poor,
  ``Application of non-orthogonal multiple access in {LTE} and {5G} networks,''
  {\em CoRR}, vol.~abs/1511.08610, 2015.

\bibitem{Ernest11}
E.~Schimmerling, {\em A Course on Set Theory}.
\newblock Cambridge Univ. Press, 2011.

\bibitem{Luo06june}
N.~Sidiropoulos, T.~Davidson, and Z.-Q. Luo, ``Transmit beamforming for
  physical-layer multicasting,'' {\em IEEE Trans. Signal Process.}, vol.~54,
  pp.~2239--2251, Jun. 2006.

\bibitem{Choi15}
J.~Choi, ``Minimum power multicast beamforming with superposition coding for
  multiresolution broadcast and application to {NOMA} systems,'' {\em IEEE
  Trans. Commun.}, vol.~63, pp.~791--800, Mar. 2015.

\bibitem{Liang07oct}
Y.~Liang, V.~Veeravalli, and H.~Poor, ``Resource allocation for wireless fading
  relay channels: {M}ax-{M}in solution,'' {\em IEEE Trans. Inf. Theory},
  vol.~53, pp.~3432--3453, Oct. 2007.

\bibitem{Scharnhorst01}
K.~Scharnhorst, ``Angles in complex vector spaces,'' {\em Acta Applicandae
  Mathematica}, vol.~69, pp.~95--103, Oct. 2001.

\bibitem{Tse05book}
D.~N.~C. Tse and P.~Viswanath, {\em Fundamentals of Wireless Communications}.
\newblock Cambridge, U.K.: Cambridge Univ. Press, 2005.

\bibitem{Luo08tsp}
E.~Karipidis, N.~Sidiropoulos, and Z.-Q. Luo, ``Quality of service and max-min
  fair transmit beamforming to multiple cochannel multicast groups,'' {\em IEEE
  Trans. Signal Process.}, vol.~56, pp.~1268--1279, Mar. 2008.

\bibitem{Ottersten14}
D.~Christopoulos, S.~Chatzinotas, and B.~Ottersten, ``Weighted fair multicast
  multigroup beamforming under per-antenna power constraints,'' {\em IEEE
  Trans. Signal Process.}, vol.~62, pp.~5132--5142, Oct. 2014.

\bibitem{Loyka01}
S.~Loyka, ``Channel capacity of {MIMO} architecture using the exponential
  correlation matrix,'' {\em IEEE Commun. Lett.}, vol.~5, pp.~369--371, Sep.
  2001.

\bibitem{Armour12}
P.~Patcharamaneepakorn, S.~Armour, and A.~Doufexi, ``On the equivalence between
  {SLNR} and {MMSE} precoding schemes with single-antenna receivers,'' {\em
  IEEE Commun. Lett.}, vol.~16, pp.~1034--1037, Jul. 2012.

\end{thebibliography}
\normalsize
\vspace{-40pt}
\begin{IEEEbiography}[{\includegraphics[width=1in,height=1.25in,clip,keepaspectratio]{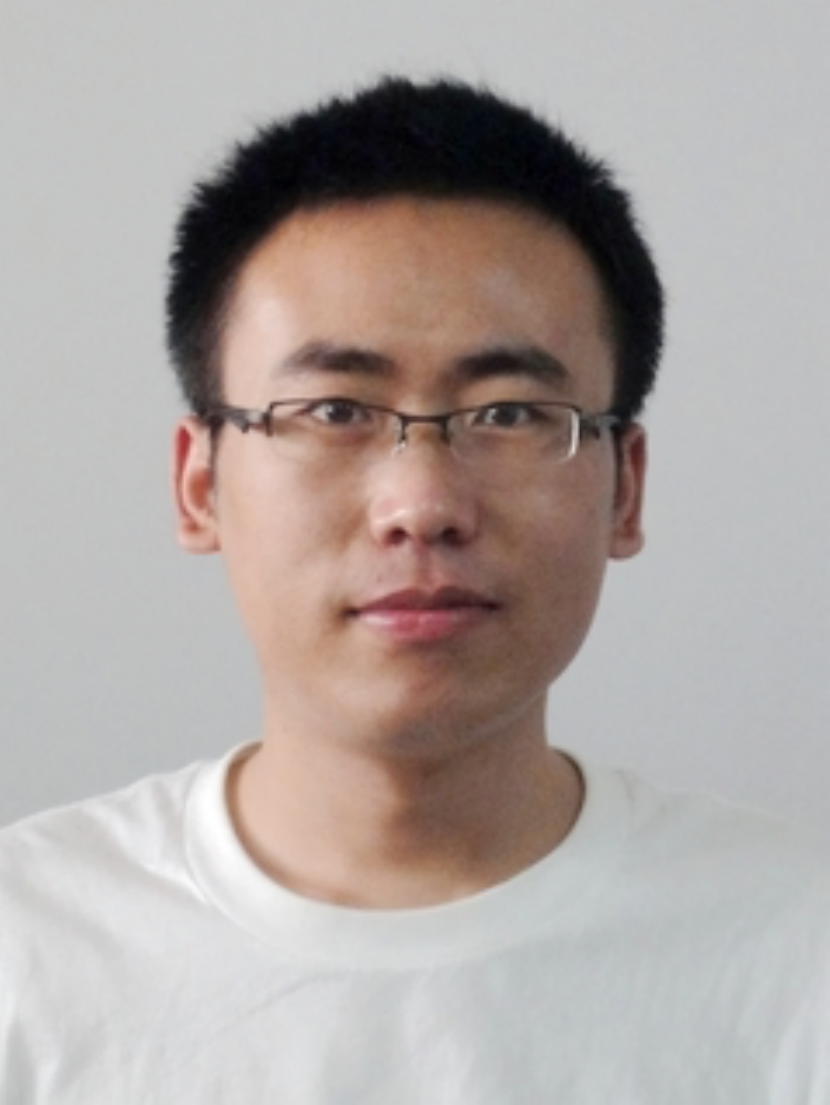}}]{Zheng Dong}
received his B.Sc. and M.Eng. degrees from the School of Information Science and Engineering, Shandong University, Jinan, China, in 2009 and 2012, respectively. Currently, he is a Ph.D. Candidate with the Department of Electronic and Computer Engineering, McMaster University, ON, Canada. His research interests include signal processing and information theory.
\end{IEEEbiography}
\vspace{-45pt}
\begin{IEEEbiography}[{\includegraphics[width=1in,height=1.25in,clip,keepaspectratio]{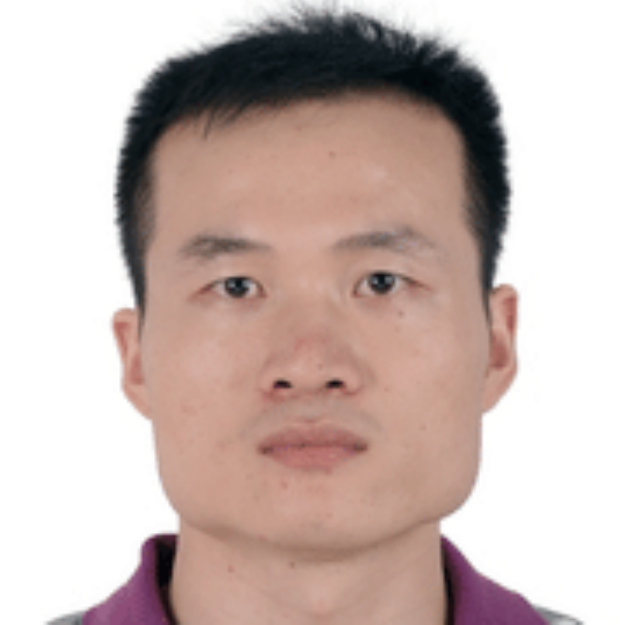}}]{Yan-Yu Zhang}
	received the B.S. degree
in Communication Engineering, the M.S. degree
in Communication and Information System and the
Ph.D. degree in Information and communication engineering from National Digital
Switching System Engineering and Technological
Research Center (NDSC), Zhengzhou, China, in 2009, 2012, and 2016, respectively. He 
is currently Lecturer at NDSC of China. His current research interests are in
 energy-efficient signal designs for radio frequency and optical wireless 
communication systems.
\end{IEEEbiography}
\vspace{-35pt}
\begin{IEEEbiography}[{\includegraphics[width=1in,height=1.25in,clip,keepaspectratio]{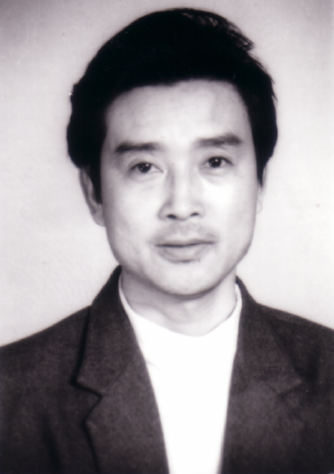}}]{Jian-Kang Zhang}
received the B.S degree in Information Science (Mathematics) from Shaanxi Normal University, M.S. degree in Information and Computational Science (Mathematics) from Northwest University, and the Ph.D. degree in Electrical Engineering from Xidian University, all in Xi'an, China. He is now Associate Professor in the Department of Electrical and Computer Engineering at McMaster University. He has held research positions at Harvard University and McMaster University. He is the co-author of the paper which received the “IEEE Signal Processing Society Best Young Author Award” in 2008.  He is currently serving as an Associate Editor for IEEE Transactions on Signal Processing and the Journal of Electrical and Computer Engineering. He has served as an Associate Editor for IEEE Signal Processing Letters.
\end{IEEEbiography}
\vspace{-35pt}

\begin{IEEEbiography}[{\includegraphics[width=1in,height=1.25in,clip,keepaspectratio]{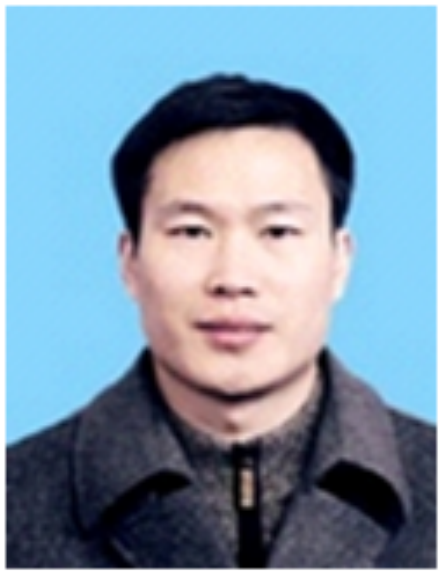}}]{Xiang-Chuan Gao}
received his B.Sc. and M.Eng. degrees from  Zhengzhou University, Zhengzhou, China, in 2005 and 2008, respectively. He then obtained his Ph.D. degree from Beijing University of Posts and Telecommunications (BUPT), Beijing, China, in 2011. He is currently a visiting professor with McMaster University, ON, Canada. His research interests include Massive MIMO, Cooperative Communications, and Visible Light Communication.
\end{IEEEbiography}


\end{document}